\documentclass[preprints]{mdpi} 
\nolinenumbers
\firstpage{1} 
\pubvolume{1}
\issuenum{1}
\articlenumber{0}
\pubyear{2021}
\copyrightyear{2020}
\datereceived{} 
\dateaccepted{} 
\datepublished{} 
\hreflink{https://doi.org/} 

\usepackage{graphicx}
\usepackage{caption}
\usepackage{subcaption}
\usepackage{amsmath}
\usepackage{textcomp}
\usepackage{xcolor}
\usepackage{soul}
\usepackage{color}

\usepackage{booktabs}

\Title{Non-binary Snow Index for Multi-Component Surfaces}

\TitleCitation{Non-binary Snow Index for Multi-Component Surfaces}

\Author{Mario M. Arreola-Esquivel $^{1}$, Carina Toxqui-Quitl $^{1,}$*, Maricela Delgadillo-Herrera $^{1}$, Alfonso Padilla-Vivanco $^{1,}$, José G. Ortega-Mendoza $^{1,}$, and Anna Carbone $^{2}$}

\AuthorNames{Mario M. Arreola-Esquivel, Carina Toxqui-Quitl, Maricela Delgadillo-Herrera, Alfonso Padilla-Vivanco, José G. Ortega-Mendoza and Anna Carbone}

\AuthorCitation{Arreola-Esquivel, M. M.; Toxqui-Quitl, C.; Delgadillo-Herrera, M.; Padilla-Vivanco, A.; Ortega-Mendoza, J. G.; Carbone, A.}

\address{%
$^{1}$ \quad Computer Vision Laboratory, Universidad Politécnica de Tulancingo, 43625, Hidalgo, México; carina.toxqui@upt.edu.mx\\
$^{2}$ \quad Department of Applied Science and Technology, Politecnico di Torino, Corso Duca degli Abruzzi 24, I-10129 Torino, Italy; anna.carbone@polito.it \\
$^{*}$ \quad Correspondence: carina.toxqui@upt.edu.mx}

\abstract{A Non-Binary Snow Index for Multi-Component Surfaces (NBSI-MS) is proposed to map snow/ice cover. The NBSI-MS is based on the spectral characteristics of different Land Cover Types (LCTs) such as snow, water, vegetation, bare land, impervious, and shadow surfaces. This index can increase the separability between  NBSI-MS values corresponding to snow from other LCTs and accurately delineate the snow/ice cover in non-binary maps. To test the robustness of the NBSI-MS, Greenland and France-Italy regions were examined where snow interacts with  highly diversified geographical ecosystem. Data recorded by Landsat 5 TM, Landsat 8 OLI, and Sentinel-2A MSI satellites have been used. The NBSI-MS performance was also compared against the well-known NDSI, NDSII-1, S3, and SWI methods and evaluated based on Ground Reference Test Pixels (GRTPs) over non-binarized results. The results show that the NBSI-MS achieves overall accuracy (OA)  ranging from 0.99 to 1 with kappa coefficient values in the same range as OA. The precision assessment confirms the performance superiority of the proposed NBSI-MS method for removing water and shadow surfaces over the compared relevant indices.}


\begin{document}
\nolinenumbers
\section{Introduction} \label{sec:one}

The cryosphere, such as snow and ice covers, reflects electromagnetic radiation that strikes the earth's surface from the sun, reducing global warming at medium and large scale \cite{ref-book1, ref-Journal1} and contributing to the water supply cycle \cite{ref-Journal2, ref-Journal3}, among other advantages.  Due to the hostile terrain conditions in cryosphere areas, Remote Sensing (RS) satellite technology  conveniently complement manual field-based techniques.  The advantage of using remote sensing derives from the  spectral and temporal resolution of the images as well as the extension of the area they cover, thus providing information relevant to modelling, further validated by means of the field data set  \cite{ref-Journal4, ref-Journal5}. Hence, the study of the cryosphere has been conducted by using RS from space since the mid-1960s \cite{ref-Journal6}. The RS provides snow/ice cover data for long-term analysis, but the lack of satellite records can limit the use of a single MultiSpectral Satellite Database (MSDB). This limitation could be resolved by combining Landsat-8 and Sentinel-2 MSDBs to provide snow/ice cover information with 2.9 days global median average revisit interval \cite{ref-Journal7, ref-Journal8,ref-Journal9}.
\par
Multiple investigations have been conducted for snow/ice cover mapping using MSDBs and several snow index methods have been applied for quantifying and categorizing the relevant information \cite{ref-Journal3, ref-Journal10, ref-proceeding1, ref-Journal11}. Snow Index-Based Methods (SIBMs) have proven to be effective as snow/ice cover extraction procedures due to their simplicity and low-cost implementation \cite{ref-Journal3, ref-Journal12}. These methods are based on an algebraic combination of spectral bands for increasing the intensity contrast between snow and non-snow pixels. Among the SIBMs existing in the literature, the Normalized Difference Snow Index (NDSI) \cite{ref-Journal12}, the S3 Index \cite{ref-Journal13}, the Normalized Difference Snow and Ice Index (NDSII) \cite{ref-Journal14}, and the Snow Water Index (SWI) \cite{ref-Journal15} are the most commonly used.
\par
Hall et al. \cite{ref-Journal12} introduced  the NDSI method for snow-cover mapping in 1995. This index is based on the snow property of  reflecting mainly visible light and absorbing radiation at infrared wavelengths; it operates via normalized differences by using the green and Near-Infrared (NIR) bands. Saito et al. \cite{ref-Journal13} introduced the S3 index and proved its better accuracy over the NDSI  \cite{ref-Journal16}. The S3 index uses the combination of Red, NIR, and Shortwave-Infrared (SWIR) bands. It has proven efficacy for snow-cover mapping under high vegetation conditions \cite{ref-Journal16}. Xiao et al.  introduced the NDSII approach in 2001, following a  methodology similar to NDSI, but with the green band replaced  by the red band \cite{ref-Journal14}. The NDSI and NDSII methods produce similar results when the Landsat Thematic Mapper (TM) data are used \cite{ref-Journal14}. The NDSI, NDSII, and S3 methods proved significant separability between snow from ice and vegetation covers  \cite{ref-Journal17, ref-Journal18, ref-Journal19}. Nevertheless, several studies \cite{ref-Journal5, ref-Journal15, ref-Journal20} have proved that these indices still classify dark forests and water as snow covers thus requiring additional masking techniques to remove spurious results from the snow-cover map \cite{ref-Journal5, ref-Journal15, ref-Journal21, ref-Journal22}. In 2019 Dixit et al. \cite{ref-Journal15} introduced the SWI technique to increase the contrast between snow/ice and other LCTs, including cloud, debris, vegetation, and water bodies. The SWI, which uses the combination of green, NIR, and SWIR bands,  has shown better results over NDSI, NDSII, and S3 methods \cite{ref-Journal15}.
\par
To remove the effect of non-snow pixels from the pure snow pixels, the NDSI, NDSII, S3, and SWI methods require an Optimal Threshold Value (OTHV) to binarize their snow map.  However, to identify the Optimal Threshold Value is a  challenge as it should be pixel-dependent across the scene  and thus  would require to be defined locally. A fixed threshold can lead to large uncertainties in the resulting snow-cover outcomes at the local scale. For snow-cover mapping based on higher-resolution imagery, a threshold value variable in space and time would be needed to improve the snow-cover map quality \cite{ref-Journal23,ref-proceeding2,carbone2010snow,valdiviezo2014hurst, ref-Journal24}.  
\par
\begin{figure}[H]
\includegraphics[width=13 cm]{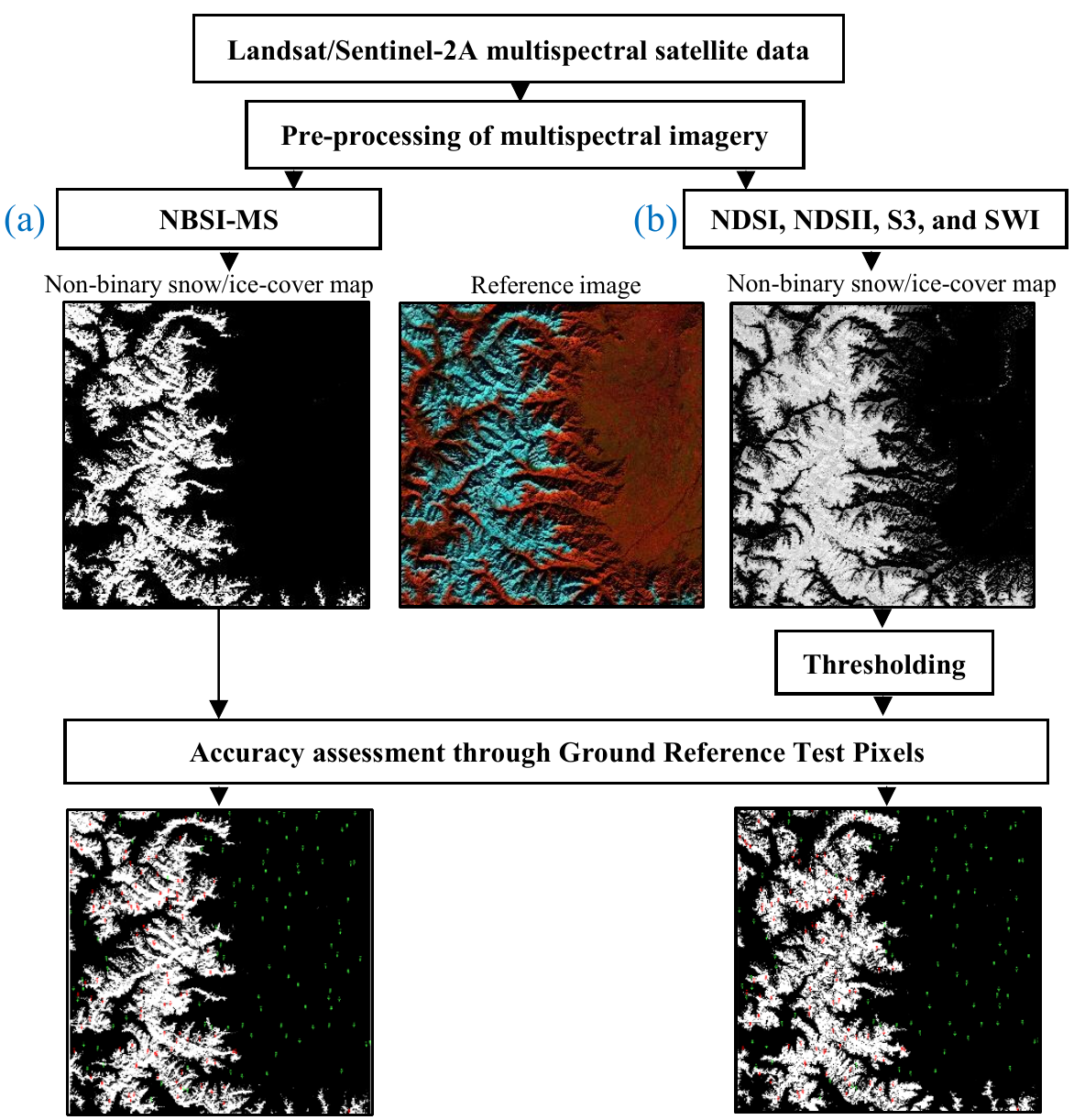}
\caption{Comparison of workflow for (a)  Non-Binary Snow Index for Multi-Component Surfaces (NBSI-MS) and (b) standard Snow Index-Based (NDSI, S3, NDSII, and SWI) methods.  (a) The proposed NBSI-MS  does not require image thresholding to enhance quality snow/ice cover maps, as opposed to (b) standard Snow Index-Based Methods proposed in the literature: NDSI, S3, NDSII, and SWI require image thresholding analysis to improve the delineation of snow-cover maps \cite{ref-Journal15, ref-Journal23,ref-Journal24, ref-Journal25}. The reference image is the false-color composite image of the region by using the combination of green, NIR, and SWIR-2 bands. It is used to visualize in blue the amount of snow cover in the image.\label{fig1}}
\end{figure}

In this paper, to the purpose of overcoming the several limitations of the existing indices, the NBSI-MS method is proposed based on the spectral characteristics of different Land Cover Types (LCTs) such as snow/ice, water, vegetation, bare land, impervious, and shadow surfaces. The NBSI-MS method uses six spectral bands Blue, Green, Red, NIR, SWIR1, and SWIR2, to increase the intensity contrast between snow and non-snow pixels. This index accurately delineates the edge of snow/ice cover without the need to set image thresholding as shown in Figure~\ref{fig1}a, using data recorded by the TM, the Operational Land Imager (OLI), and the MultiSpectral Instrument (MSI) satellite sensors.
\par
The snow-cover mapping performance of the proposed NBSI-MS is compared against NDSI, NDSII, S3, and SWI methods in the presence of water, vegetation, bare land, impervious, and shadow surfaces. The snow/ice cover maps produced by all indices are evaluated in Greenland and France-Italy regions in the same image conditions according to the pre-processing steps and in non-binarized results as follows: a) qualitatively through visual inspection and b) quantitatively based on GRTPs validation data for the precision assessment. The very good  agreement between qualitative and quantitative results confirms superiority of the NBSI-MS method to reject the water and shadow surfaces correctly, whereas the NDSI, NDSII, S3, and SWI fail to suppress them.
\par
\par
The remainder of this paper is organized as follows: Section \ref{sec:two} contains the description of the selected test areas, data collection, image pre-processing, and application of the Snow-Cover Indices (SCIs): NDSI, NDSII, S3, SWI, and the proposed NBSI-MS methods. Section \ref{sec:three} reports the results of the SCIs evaluation as follows: a) qualitatively through visual inspection and b) quantitatively based on GRTPs in non-binarized snow-cover maps for the precision assessment. In Section \ref{sec:four}, the most outstanding results are discussed. Finally, Section \ref{sec:five} reports our main conclusions.  
\section{Materials and Methods}
\label{sec:two}
Digital image processing has been performed by using  ENVI v5.3 (Exelis Visual Information Solutions, Boulder, CO, USA), Sen2Cor algorithm on the SentiNel Application Platform (SNAP) created by the European Space Agency (ESA), MATLAB R2019a, ArcGIS 10.5 (Environmental System Research Institute California, CA, USA), and Google Earth Pro$^{TM}$ (Google Inc; Menlo Park, CA, USA) software packages.
\subsection{Test Areas}
\begin{figure}[H]
\includegraphics[width=13 cm]{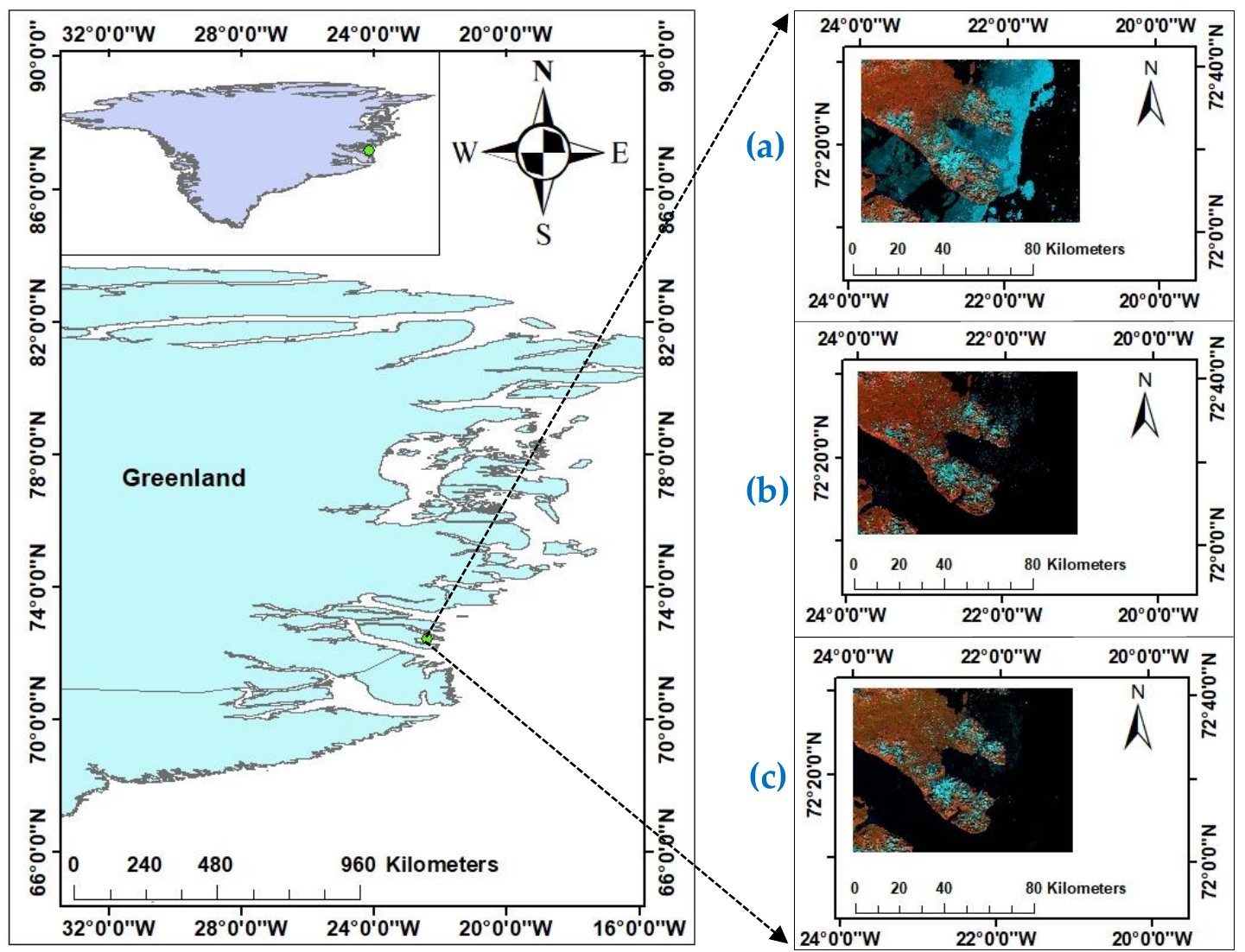}
\caption{Location of the  Greenland area under study and the false-color composite images used to visualize the amount of the snow cover. The left panel represents the precise location of the area, while the right panel shows  the false-color composite images using the green, NIR, and SWIR-2 bands of this region for (a) Landsat 5 TM, (b) Landsat 8 OLI, and (c) Sentinel-2A MSI.
\label{fig2}}
\end{figure}

 A region of Greenland covered with snow/ice, bare land, and a large water surface is analyzed to evaluate the SCIs performance.  Figure~\ref{fig2} shows the details of this area having a spatial extent of 97.95$\times$72.99 km recorded by Landsat and Sentinel-2A, with  center located  at 72$^{\circ}$22'15.01" N, 22$^{\circ}$28'35.85" W.
 \par
\begin{figure}[H]
\includegraphics[width=13 cm]{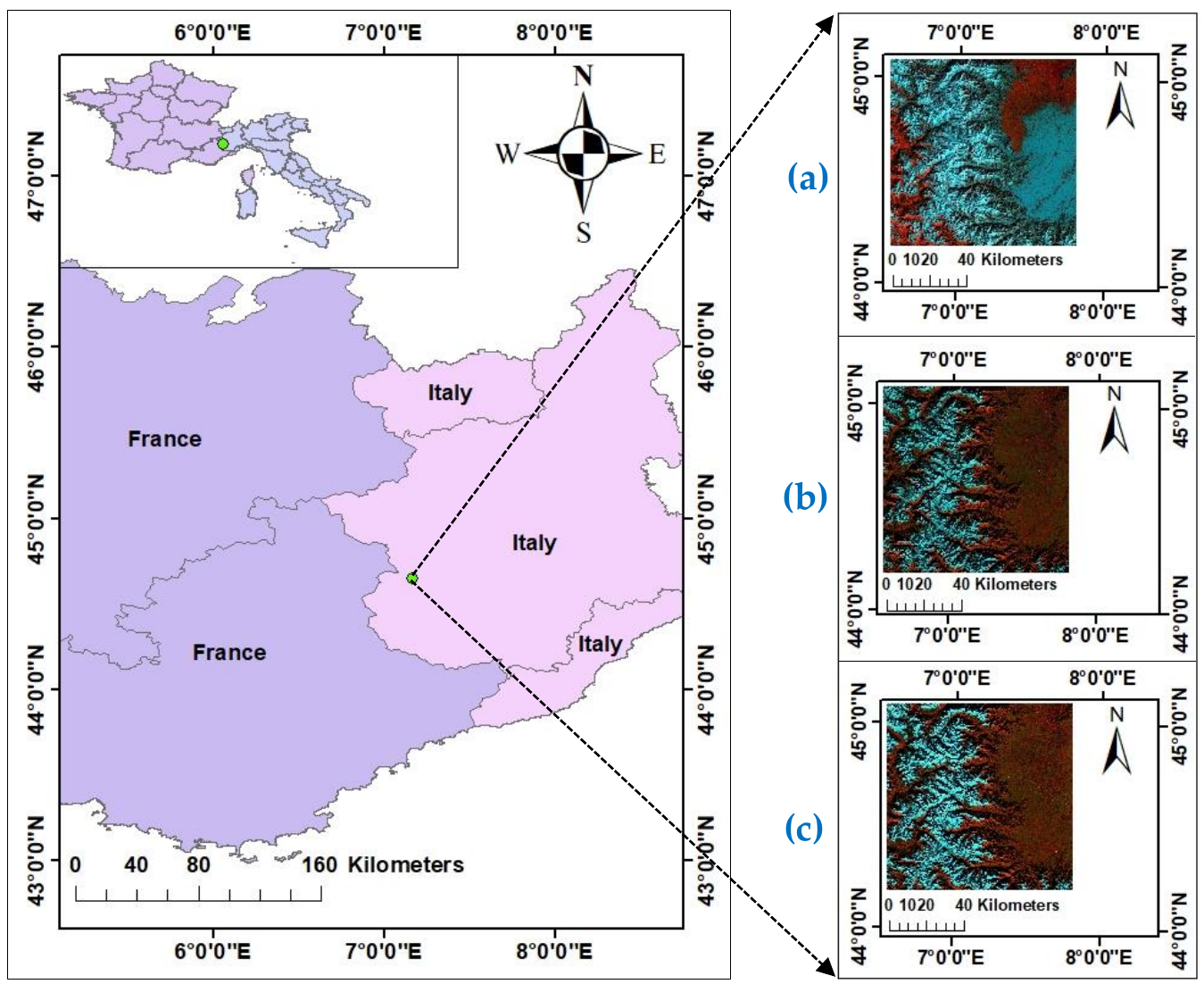}
\caption{Location of the  France-Italy region under study and false-color composite images of the region. The left panel represents the precise location of the study region between Italy and France, while the right panel shows the composite images using the green, NIR, and SWIR-2 bands  for (a) Landsat 5 TM, (b) Landsat 8 OLI, and (c) Sentinel-2A MSI.
\label{fig3}}
\end{figure}

Additionally, a region located at the France-Italy border is selected to evaluate all indices under highly diverse geographical conditions. The region's LCTs include water, vegetation, bare land, urban areas, hilly land, and snow/ice. This scene also includes Hilly Shadow over Vegetation (HS-V) and Hilly Shadow over Bare Land (HS-BL). With such LCTs diversity, it is challenging for the SCIs to map the snow/ice cover in this scene. The area is characterized by  a spatial extent of 100.8$\times$101.01 km and is centered at the coordinates 44$^{\circ}$38'40.65" N, 7$^{\circ}$9'57.30" E (see Figure~\ref{fig3}).

\subsection{Data}
The Landsat and Sentinel-2 datasets were downloaded from the United States Geological Survey (USGS) web portal (https://earthexplorer.usgs.gov/)  \cite{ref-url1}. The Landsat Level 1 Terrain Corrected (L1T) and Sentinel-2 Level 1C (L1C) data are in the Universal Transverse Mercator (UTM) map projection. Table~\ref{tab1} shows the specifications of the Landsat and Sentinel-2A datasets used in this research. Greenland and France-Italy test areas, recorded by Landsat 8 OLI and Sentinel-2A MSI  in the same period of the year to have a comparable snow/ice cover in the scenes, were examined. However, the datasets recorded by the TM sensor refer to a different period than OLI and MSI sensors as the Landsat 5 TM ceased in November 2011 \cite{ref-url2}. Nonetheless, since Landsat 5 TM collected  data continuously for nearly 29 years, it still constitutes an invaluable source for long-term RS analysis. Additionally, this study considered the percentage of snow/ice cover   difference in the test areas (see section \ref{sec:sec:two} for precision assessment) registered by the TM sensor from those of the OLI and MSI sensors as reported in 5th column of  Table 1. Since the SCIs can remove the cloud-contaminated pixels partially but not fully, the images of the three sensors with minimum cloud coverage were chosen on the basis of the cloud percentages as provided by each dataset.  
\newpage
\end{paracol}
\nolinenumbers
\nointerlineskip

\begin{specialtable}[H]
\caption{Summary of relevant data for Landsat and Sentinel-2A imagery in Greenland and France-Italy regions.  \label{tab1}}
\centering
\begin{tabular}{lcccccc}
  \toprule
\multicolumn{1}{c}{\textbf{Region}} & \textbf{Satellite} & \textbf{Acquisition Date}                                    & \textbf{\begin{tabular}[c]{@{}c@{}}Cloud \\ cover \end{tabular}} &  \textbf{\begin{tabular}[c]{@{}c@{}}Snow \\ cover \end{tabular}} & \textbf{\begin{tabular}[c]{@{}c@{}}Sun zenith \\ angle\end{tabular}} & \textbf{\begin{tabular}[c]{@{}c@{}}Sun azimuth \\ angle\end{tabular}} \\ \hline
\textbf{Greenland}                    & Landsat 5 TM       & July 11, 1991                                                & 2.00 \%              & 18.02\%             & 50.44°                                                               & 166.83°                                                               \\ \hline
\textbf{}                             & Landsat 8 OLI      & July 22, 2016                                                & 1.09 \%              & 5.09\%              & 52.11°                                                               & 177.72°                                                               \\ \hline
\textbf{}                             & Sentinel-2A MSI    & July 21, 2016                                                & 0 \%                 & 5.09\%              & 52.2°                                                                & 181.8°                                                                \\ \hline
\textbf{France-Italy}                 & Landsat 5 TM       & March 10, 1993                                               & 1.00\%               & 37.91\%             & 55.9°                                                                & 141.85°                                                               \\ \hline
\textbf{}                             & Landsat 8 OLI      & \begin{tabular}[c]{@{}c@{}}December 06, \\ 2016\end{tabular} & 1.11\%               & 25.86\%             & 68.84°                                                               & 163.67°                                                               \\ \hline
                                      & Sentinel-2A MSI    & \begin{tabular}[c]{@{}c@{}}December 01, \\ 2016\end{tabular} & 0.27\%               & 25.86\%             & 67.23°                                                               & 169.44°           \\                                               
  \bottomrule
  \end{tabular}
\end{specialtable}

\begin{paracol}{2}
\nolinenumbers
\switchcolumn

Since different satellite sensors are used in this study, a comparison of the spatial and spectral resolutions must be considered \cite{ref-Journal26}. Table~\ref{tab2} shows the specifications of the selected bands for the satellite sensors used in this research. To perform consistent algebraic operations,  Sentinel-2A SWIR bands were resized to the same spatial resolution (10 m) by using the nearest neighborhood algorithm approach on SNAP \cite{ref-Journal27} . 
\end{paracol}
\nointerlineskip

\begin{specialtable}[H]
\caption{Wavelength and spatial resolution specifications of the bands used for estimating the Snow Cover Indexes  for Landsat 5 TM, Landsat 8 OLI and Sentinel-2A imagery \cite{ref-url3,ref-url4}. \footnotesize{Note: The spatial resolution specifications are given for a pixel.} \label{tab2}}
\centering
  \begin{tabular}{ccccccc}
    \toprule
    \multirow{3}{*}{\textbf{BANDS}} &
      \multicolumn{2}{c }{\textbf{Landsat 5 TM}} &
      \multicolumn{2}{c }{\textbf{Landsat 8 OLI}} &
      \multicolumn{2}{c }{\textbf{Sentinel-2 MSI}} \\ \cmidrule{2-7}
      &\scriptsize { { Wavelength}} & \scriptsize {{Spatial}} & \scriptsize {{ Wavelength}} & \scriptsize {{Spatial}} & \scriptsize {{ Wavelength}} & \scriptsize {{Spatial}} \\
      & \scriptsize {{$(\mu m)$}} & \scriptsize {{resolution $(m)$}} & \scriptsize {{$(\mu m)$}} & \scriptsize {{resolution $(m)$}} & \scriptsize {{$(\mu m)$}} & \scriptsize {{resolution $(m)$}} \\
      \midrule
\textbf{\footnotesize {Blue $(B_{B})$}} & 0.45 - 0.52 & 30 & 0.45-0.51 & 30 & 0.46-0.52 & 10 \\
\textbf{\footnotesize{Green $(B_{G})$}} & 0.52 - 0.60 & 30 & 0.53-0.59 & 30 & 0.55-0.58 & 10 \\
\textbf{\footnotesize{Red $(B_{R})$}} & 0.63 - 0.69 & 30 & 0.64-0.67 & 30 & 0.64-0.67 & 10 \\
\textbf{\footnotesize{NIR $(B_{NIR})$}} & 0.76 - 0.90 & 30 & 0.85-0.88 & 30 & 0.78-0.90 & 10 \\
\textbf{\footnotesize{SWIR-1 $(B_{SWIR-1})$}} & 1.55 - 1.75 & 30 & 1.57-1.65 & 30 & 1.57-1.65 & 20 \\
\textbf{\footnotesize{SWIR-2 $(B_{SWIR-2})$}} & 2.08 - 2.35 & 30 & 2.11-2.29 & 30 & 2.10-2.28 & 20 \\

   \bottomrule
  \end{tabular}
\end{specialtable}

\begin{paracol}{2}
\switchcolumn
\subsection{Multispectral image pre-processing}

Pre-processing steps must be taken into account to reduce sensor, solar, diffraction \cite{Urcid2005}, and atmospheric effects \cite{Olvera2020, GonzalezAmador2020}  among satellite data recorded on different dates  \cite{ref-Journal28, ref-Journal29}.  Figure~\ref{fig4} shows the pre-processing workflow for the Landsat and Sentinel-2A datasets used in this study. The Landsat satellite data were pre-processed in ENVI, employing radiometric calibration tools using the Landsat header MTL metadata file information to obtain the Top-Of-Atmosphere (TOA) reflectance data. On the other hand, the Sentinel-2A L1C data were pre-processed using the Sen2Cor algorithm on the SNAP to produced Bottom-Of-Atmosphere (BOA) reflectance data. According to Merzah et al. \cite{ref-proceeding3}, the Internal Average Relative Reflectance (IARR) correction technique, removes or reduces the effects of the atmosphere providing a clean spectral curve to be interpreted. In this research, the IAR reflectance correction was applied to the Landsat and Sentinel-2A datasets to reduce the atmosphere effects and obtain similar atmospheric conditions \cite{ref-Journal30, ref-Journal31, ref-book2}. 

\begin{figure}[H]
\includegraphics[width=13 cm]{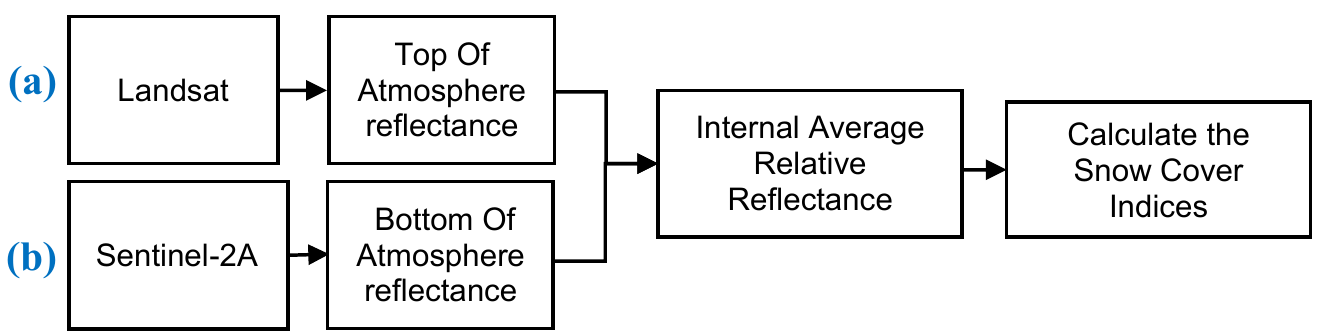}
\caption{Pre-processing steps applied to the satellite data before the index calculation: (a) Landsat and (b) Sentinel-2A. \label{fig4}}
\end{figure}
\subsection{Methodology}

The proposed NBSI-MS snow cover delineation index performance is compared against the well-known SIBMs:  NDSI \cite{ref-Journal12}, S3 \cite{ref-Journal13}, NDSII \cite{ref-Journal14}, and SWI \cite{ref-Journal15}. The selected bands to calculate all indices are specified in Table~\ref{tab2}, and the mathematical expressions of NDSI, NDSII, S3, and SWI are shown in Table~\ref{tab3}.
According to Stevens’s taxonomy of measurement scale (nominal, ordinal,interval,  and  ratio) \cite{Stevens1946Onthetheory}, the  SCI  index,  including  the  proposed  one,  can be  considered   ordinal  measurements as  they  permit  a  rank ordering of relative snow content and remain invariant under monotonic increasing transformations

\begin{specialtable}[H]
\centering
\caption{Summary of the Snow Index-Based Methods (SIBMs) used in the research to compare the capacity of the Non-Binary Snow Index for Multi-Component Surfaces method. \label{tab3}}
\begin{tabular}{ccc}
\toprule
\textbf{SIBMs} & \textbf{Mathematical expression} & \textbf{Reference} \\ \hline
\textbf{NDSI}                   &   $\displaystyle \frac{B_{G}-B_{SWIR-1}}{B_{G}+B_{SWIR-1}}$                               & \cite{ref-Journal12}         \\ \hline
\textbf{S3}                     & $\displaystyle \frac{B_{NIR}(B_{R}-B_{SWIR-1})}{(B_{NIR}+B_{R})(B_{NIR}+B_{SWIR-1})}$                                                                                             & \cite{ref-Journal13}       \\ \hline
\textbf{NDSII}                  &    $\displaystyle \frac{B_{R}-B_{SWIR-1}}{B_{R}+B_{SWIR-1}}$                                                             & \cite{ref-Journal14}        \\ \hline
\textbf{SWI}                    & $\displaystyle \frac{B_{G}(B_{NIR}-B_{SWIR-1})}{(B_{G}+B_{NIR})(B_{NIR}+B_{SWIR-1})}$                                                                                                                             & \cite{ref-Journal15}       \\
\bottomrule
\end{tabular}
\end{specialtable}

\subsection{Non-Binary Snow Index for Multi-component Surfaces (NBSI-MS)}

The infrared band registered by an optical sensor captures the thermal radiation of objects in a given scene. In contrast, a visible band mainly records optical reflection information \cite{ref-Journal32}. The proposed NBSI-MS method takes advantage of six spectral bands ($B_{B}$, $B_{G}$, $B_{R}$, $B_{NIR}$, $B_{SWIR-1}$, and $B_{SWIR-2}$ defined in Table~\ref{tab2}) to achieve a maximum separability between snow and non-snow pixels. According to Gong et al. \cite{ref-Journal33}, the global land cover resides in different LCTs such as vegetation, water, impervious (e.g., urban areas, roads, industrial areas), bare land (e.g., beaches, deserts, rocks, gravel pits), and snow/ice \cite{ref-Journal34}. The NBSI-MS is based on the spectral characteristics of these global LCTs classification plus HS-V and HS-BL. These seven LCTs spectral signatures were obtained using the mean reflectance values of one hundred samples selected on each LCT over the France-Italy region. Figures~\ref{fig:fig5a} and \ref{fig:fig5b} show these spectral signatures and Table~\ref{tab4} displays its mean reflectance values using Landsat 8 OLI and Sentinel-2A MSI preprocessed data.

\newpage

\end{paracol}
\nointerlineskip

\begin{figure}[H]
\captionsetup[subfigure]{justification=centering}
  \begin{subfigure}[b]{0.5\textwidth}
    \includegraphics[width=0.8\textwidth]{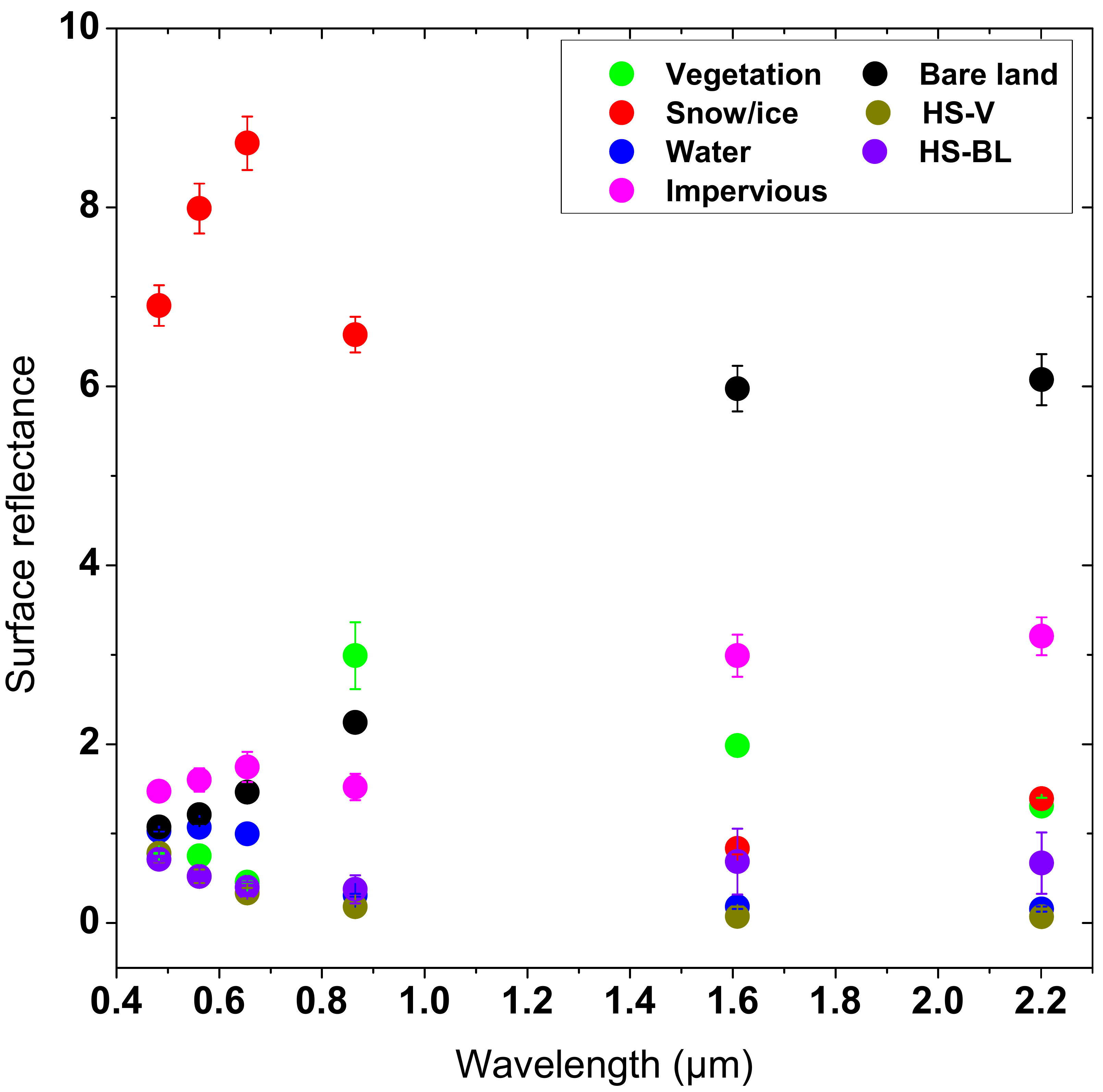}
    \caption{Landsat 8 OLI.}
    \label{fig:fig5a}
  \end{subfigure}
  \begin{subfigure}[b]{0.5\textwidth}
    \includegraphics[width=0.8\textwidth]{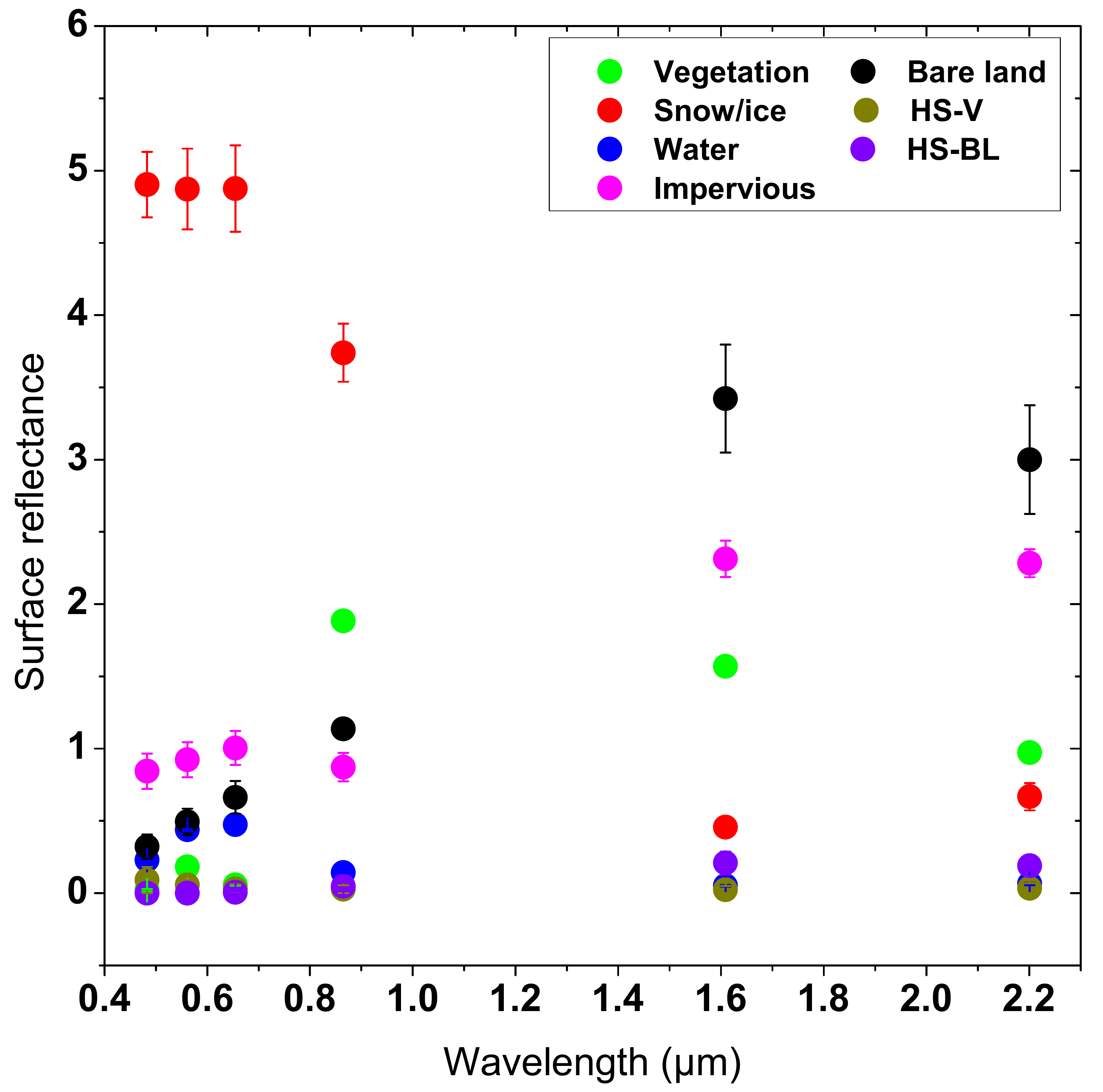}
    \caption{Sentinel-2A MSI.}
    \label{fig:fig5b}
  \end{subfigure}
  \caption{Spectral signatures (mean $\pm$ standard deviation) for (a) Landsat OLI 8 and (b) Sentinel 2A MSI. The signatures of each type were obtained using the mean of 100  pixels selected in each Land Cover Type. \footnotesize{Note: HS-BL is the Hilly Shadow over Bare Land and HS-V is the Hilly Shadow over Vegetation.}}
\end{figure}

\begin{paracol}{2}
\switchcolumn

The LCTs reflectance between Sentinel-2A MSI, Landsat 8 OLI, and Landsat 5 TM can be slightly different as these satellite technologies are similar but not identical \cite{ref-Journal26}. In spite of the LCTs reflectance differences, as shown in Table~\ref{tab4}, the proposed NBSI-MS can take advantage of the spectral bands registered by Landsat and Sentinel-2A satellite technologies.

\end{paracol}
\nointerlineskip

\begin{specialtable}[H]
\centering
\caption{Mean reflectance values of one hundred pixels used in Figures \ref{fig:fig5a}  and \ref{fig:fig5b} in the France-Italy region, where HS-BL is the Hilly Shadow over Bare Land, and HS-V is the Hilly Shadow over Vegetation. \label{tab4}}
\begin{tabular}{ccccccccc}
\toprule
\multirow{2}{*}{\textbf{Satellite}}       & \multicolumn{1}{c}{\multirow{2}{*}{\textbf{Bands}}} & \multicolumn{7}{c}{\textbf{Reflectance   values}}                                                                                                                                                                                                                                        \\ \cline{3-9} 
                                          & \multicolumn{1}{c}{}                                & \multicolumn{1}{c}{\textbf{Vegetation}} & \multicolumn{1}{c}{\textbf{Snow/ice}} & \multicolumn{1}{c}{\textbf{Water}} & \multicolumn{1}{c}{\textbf{Impervious}} & \multicolumn{1}{c}{\textbf{Bare   land}} & \multicolumn{1}{c}{\textbf{HS-V}} & \multicolumn{1}{c}{\textbf{HS-BL}} \\ \hline
\multirow{6}{*}{\textbf{Landsat 8 OLI}}   &                        $B_{B}$                                                  & 0.78                                  & 6.90                                & 1.03                             & 1.47                                  & 1.08                                   & 0.79                            & 0.71                              \\ \cline{2-9} 
                                          &           $B_{G}$                                                               & 0.75                                  & 7.99                                & 1.07                             & 1.60                                  & 1.21                                   & 0.52                            & 0.52                             \\ \cline{2-9} 
                                          &           $B_{R}$                                                              & 0.46                                  & 8.72                                & 1.00                             & 1.75                                  & 1.47                                   & 0.34                            & 0.40                             \\ \cline{2-9} 
                                          &               $B_{NIR}$                                                           & 2.99                                  & 6.58                                & 0.31                             & 1.52                                  & 2.24                                   & 0.18                            & 0.38                             \\ \cline{2-9} 
                                          &              $B_{SWIR-1}$                                                            & 1.99                                   & 0.83                                & 0.18                             & 2.99                                  & 5.97                                   & 0.08                            & 0.69                             \\ \cline{2-9} 
                                          &                  $B_{SWIR-2}$                                     & 1.31                                  & 1.39                                & 0.16                             & 3.21                                  & 6.07                                   & 0.07                            & 0.67                             \\ \hline
\multirow{6}{*}{\textbf{Sentinel-2A MSI}} &               $B_{B}$                                       & 0.02                                  & 4.90                                & 0.23                             & 0.84                                  & 0.32                                   & 0.09                            & 0.00                            \\ \cline{2-9} 
                                          &               $B_{G}$                                        & 0.18                                  & 4.87                                & 0.44                             & 0.92                                  & 0.49                                   & 0.06                            & 0.00                            \\ \cline{2-9} 
                                          &               $B_{R}$                                        & 0.06                                  & 4.88                                & 0.48                              & 1.00                                  & 0.66                                   & 0.03                            & 0.00                             \\ \cline{2-9} 
                                          &               $B_{NIR}$                                        & 1.88                                  & 3.74                                & 0.14                             & 0.87                                  & 1.14                                   & 0.03                            & 0.05                              \\ \cline{2-9} 
                                          &               $B_{SWIR-1}$                                        & 1.57                                  & 0.46                                & 0.056                             & 2.316                                   & 3.42                                    & 0.02                            & 0.21                             \\ \cline{2-9} 
                                          &               $B_{SWIR-2}$                                        & 0.97                                  & 0.67                                & 0.06                             & 2.28                                  & 3.00                                   & 0.03                            & 0.19                             \\ 
\bottomrule
\end{tabular}
\end{specialtable}                                           

\begin{figure}[h!]
\captionsetup[subfigure]{justification=centering}
  \begin{subfigure}[b]{0.5\textwidth}
    \includegraphics[width=0.8\textwidth]{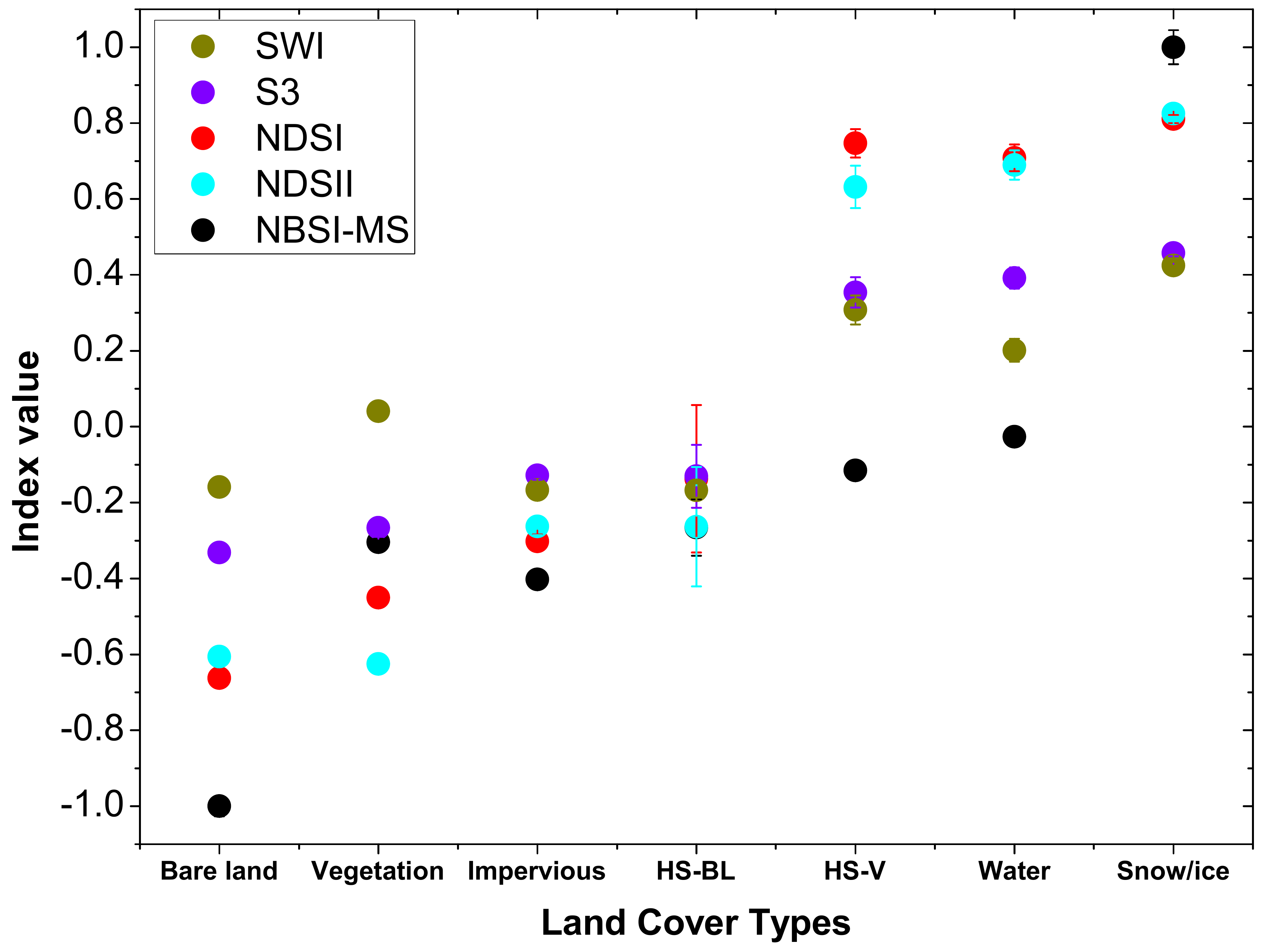}
    \caption{Landsat 8 OLI.}
    \label{fig:fig6a}
  \end{subfigure}
  \begin{subfigure}[b]{0.5\textwidth}
    \includegraphics[width=0.8 \textwidth]{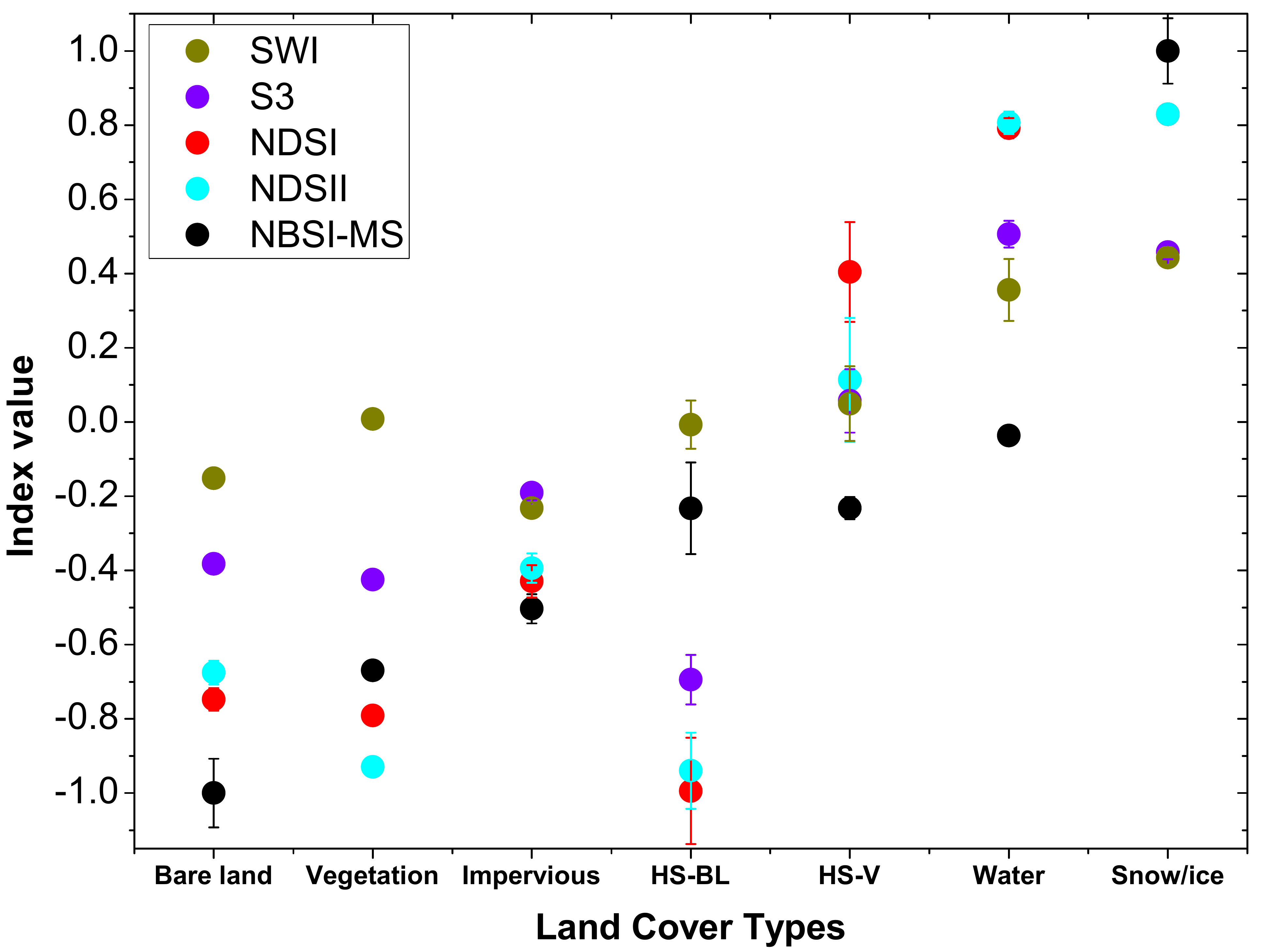}
    \caption{Sentinel-2A MSI.}
    \label{fig:fig6b}
  \end{subfigure}
  \caption{Plots of the resulting index values (mean $\pm$ standard deviation) obtained by computing the mathematical expressions in Table \ref{tab3} and Eq. (\ref{eqn:eqn1}) by taking the values in Table \ref{tab4} for the different Snow-Cover Indices (SCIs). The list of SCIs is given as the snow/ice index values increase. The NBSI-MS actual values were normalized to analyze the different SCIs under the same scale while HS-BL is the Hilly Shadow over Bare Land, and HS-V is the Hilly Shadow over Vegetation.}
\end{figure}

\begin{paracol}{2}
\switchcolumn

Typically, in the visible and NIR wavelengths, snow is brighter than vegetation, water, impervious, bare land, HS-V, and HS-BL \cite{ref-Journal16, ref-Journal34}, as shown in Figures~\ref{fig:fig5a} and \ref{fig:fig5b}. The NBSI-MS method takes advantage of these snow/ice spectral characteristics by adding $B_{G}$, $B_{R}$, and $B_{NIR}$ bands, to increase the separability of pure snow/ice surface from other LCTs. The proposed NBSI-MS expression is given by:  

\begin{equation}
\label{eqn:eqn1}
NBSI-MS = k(B_{G}+B_{R}+B_{NIR})-\left( \frac{B_{B}+B_{SWIR-2}}{B_{G}}+B_{SWIR-1}\right).
\end{equation}

For the Landsat imagery, each band must be multiplied by its assigned reflectance value allocated in the MTL metadata file. Fixed empirical coefficients can be determined based on examining reflectance properties of different LCTs as in \cite{ref-Journal35}. {\em k} = 0.36 is a fixed empirical coefficient used to increase the intensity contrast between snow and other LCTs pixels. {\em k} was determined by using an iterative process to identify the parameter [0.34 to 0.38] that maximizes the separability of snow and non-snow surfaces. It was found that {\em k} = 0.36 enhance separability by forcing snow pixels above 0 and other LCTs pixels below 0 to stabilize non-binarization maps. 
\par
The subtracting term of the NBSI-MS expression is called the LCTs mask. It allows to suppress the six LCTs apart from snow/ice with high performance.  
The effective operation of the NBSI-MS method relies on the LCTs mask, which uses the $B_{SWIR-2}$ that is the second band with the lowest reflectance across the snow spectral signature. At the same time, this band is highly reflected by bare land, impervious vegetation, and HS-BL, as shown in Figures \ref{fig:fig5a} and \ref{fig:fig5b}. Furthermore, $B_{SWIR-2}$ is added to $B_{B}$ to obtain a high reflectance in the LCTs except for the snow. The $B_{G}$ was selected to balance the values resulting from $B_{B} + B_{SWIR-2}$ in the snow areas. Thus, the term $({B_{B} + B_{SWIR-2})}/{B_{G}}$ tends to take values close to 1 for snow covers, while for the background, higher values than 1. Finally, the $B_{SWIR-1}$ is added at the end of the mask by the same criteria as  $B_{SWIR-2}$. In this way, the NBSI-MS method can effectively remove the background with negative values since the LCTs mask is subtracted. 
The  calculation of the proposed NBSI-MS index according to Eq. $(\ref{eqn:eqn1})$ is intended to produce the following results:

\begin{enumerate}
    \item Deep-water and impervious surfaces will be separated from snow since these LCTs reflect well blue light. Also the possible LCTs saturated pixels of  $B_{B}$  will yield negative values since they appear in the LCTs mask with the subtracting term. 
	\item Vegetation is removed because it exhibits two reflectance peaks, respectively in the green and in the NIR bands, which are lower than snow/ice reflectance. 
	\item Water and HS-V are suppressed, considering their lowest reflectance in the NIR. Their signatures appear with reflectance lower  than snow in all six bands. 
	\item Impervious and bare-land are suppressed as both show roughly flat spectral signatures at the visible and NIR bands with lower reflectance compared to snow. Both components will provide negative values because of their higher reflectance at the SWIR bands. 
\end{enumerate}
In summary, the added factors in the first parentheses of the NBSI-MS expression Eq. $(\ref{eqn:eqn1})$ tends to enhance the snow/ice cover mapping, while the LCTs mask term tends to suppress the contribution from vegetation, water, impervious, bare land, HS-V, and HS-BL.

The rule to evaluate the NBSI-MS maps about the snow or not snow pixels classification is:

\begin{equation} \label{eq:2}
  p(x,y) =
  \begin{cases}
                                   1 & \text{if NBSI-MS $>$ 0} \\
                                   0 & \text{Otherwise.} \\
  \end{cases}
\end{equation}
where {\em x,y} is the position of the reference test pixels, {\em p(x,y)}   are the  NBSI-MS index values according to Eq. (\ref{eqn:eqn1}), $p(x,y)=1$ and  $p(x,y)=0$  correspond respectively to snow and non-snow. 
 For the sake of completeness, we note that the snow not-snow rule described above reduces the range of values and information carried by the several bands involved in the Eq. (\ref{eqn:eqn1}). The precision should be intended as a relative assessment of the NBSI-MS index in comparison to the indexes. Further multispectral visualization of the NBSI-MS index can be envisioned and will be matter of future work.


\section{Results}\label{sec:three}
To compare the capability of the SCIs to differentiate between snow and background (land, impervious, vegetation, water, HS-V, and HS-BL) on the France-Italy region, these were computed using the LCTs mean spectral values of Table~\ref{tab4}. Figures~\ref{fig:fig6a} and \ref{fig:fig6b} depict the resulting plots from the computed SCIs via Landsat 8 OLI and Sentinel-2A MSI. The NBSI-MS values have been normalized in the range $[-1, 1]$ to compare all the indexes on the same scale. The results reveal that the proposed NBSI-MS method can highlight the snow/ice surface with better performance over other SCIs since the background is suppressed with NBSI-MS values below zero. On the other hand, the NDSI, NDSII, S3, and SWI methods cannot identify the snow because their positive index values corresponding to water and HS-V are close to the snow ones. This result suggests the need to be masked or find an optimal threshold value during snow cover delineation.
\end{paracol}
\nointerlineskip

\begin{paracol}{2}
\switchcolumn

Normalized and not-normalized NBSI-MS SCI values shown in Table~\ref{tab5} were computed using the expressions in  Table~\ref{tab3}, Eq. $(\ref{eqn:eqn1})$, and the mean spectral values of  Table~\ref{tab4}. The outcomes show that the proposed NBSI-MS method can highlight the snow/ice surface with a Maximum Positive Index Value (MPIV) of 8.10 for Landsat 8 OLI data. Large negative NBSI-MS values, ranging from -0.29 to -10.90, are found for the LCT components thus allowing to easily remove the background. Likewise, when Sentinel-2A MSI data are used, the NBSI-MS method has an MPIV of 3.26, and the background is suppressed with negative NBSI-MS values, ranging from -0.34 to -9.32. As a result, the large contrast of the NBSI-MS values between snow/ice and background shows that there will be no doubt in separating them on an NBSI-MS image. 
\newpage

\end{paracol}
\nointerlineskip

\begin{specialtable}[H]
\centering
\caption{Values of the indexes plotted in Figures \ref{fig:fig6a} and \ref{fig:fig6b} for the different Snow-Cover Indices. The NBSI-MS values were normalized as shown in parenthesis in the range [-1, 1] to compare the proposed method in the same range as NDSI, NDSII, S3, and SWI. \footnotesize{Note: The snow index value is a single number that quantifies snow cover computed by Eq. (\ref{eqn:eqn1}) and expressions in Table \ref{tab3} taking the values in Table \ref{tab4}. } \label{tab5} }
\begin{tabular}{ccccccc}
\toprule
\multirow{2}{*}{\textbf{Satellite}}                                                 & \multirow{2}{*}{\textbf{Land   cover type}} & \multicolumn{5}{c}{\textbf{Index   values}}              \\ \cline{3-7}
                                                                           &                                    & \textbf{NBSI-MS}  & \textbf{NDSI}    & \textbf{NDSII}   & \textbf{S3}      & \textbf{SWI}     \\ \hline
\multirow{7}{*}{\begin{tabular}[c]{@{}c@{}}\textbf{Landsat 8}\\  \textbf{OLI}\end{tabular}}  & \textbf{Bare land                         } & -10.99 (-1) & -0.66 & -0.61 & -0.33 & -0.16 \\ \cline{2-7}
                                                                           & \textbf{Impervious}                         &  -4.42 (-0.40)& -0.30 & -0.26 & -0.13 & -0.17 \\ \cline{2-7}
                                                                           & \textbf{Vegetation}                         & -3.35 (-0.30)  & -0.45 & -0.63 & -0.27 & 0.04  \\ \cline{2-7}
                                                                           & \textbf{Water}                              &-0.23 (-0.03) & 0.71  &  0.69  & 0.39  &  0.20  \\ \cline{2-7}
                                                                           & \textbf{HS-V}                               & -1.27 (-0.12) &  0.75  &  0.63  &  0.35  &  0.31  \\ \cline{2-7}
                                                                           & \textbf{HS-BL}                              & -2.92 (-0.27) & -0.14 & -0.26 & -0.13 & -0.17 \\ \cline{2-7}
                                                                           & \textbf{Snow}                               & \ 8.10 (1)  &  0.81  &  0.83  &  0.46  &  0.42  \\ \hline
\multirow{7}{*}{\begin{tabular}[c]{@{}c@{}}\textbf{Sentinel-2} \\ \textbf{MSI}\end{tabular}} & \textbf{Bare land                         } & -9.32 (-1) & -0.75 & -0.68 & -0.38 & -0.15  \\ \cline{2-7}
                                                                           & \textbf{Impervious}                         & -4.70 (-0.50)  & -0.43 & -0.40 & -0.19 & -0.23 \\ \cline{2-7}
                                                                           & \textbf{Vegetation}                         & -6.24 (-0.67) & -0.79 & -0.93 & -0.42 & 0.01   \\ \cline{2-7}
                                                                           & \textbf{Water}                              & -0.34 (-0.04)  & \ 0.79  &  0.81  &  0.51  &  0.36  \\ \cline{2-7}
                                                                           & \textbf{HS-V}                               & -2.16 (-0.23) & 0.40  & 0.11  & 0.06  & 0.05  \\ \cline{2-7}
                                                                           & \textbf{HS-BL}                              & -2.17 (-0.23)  & -0.99 & -0.94 & -0.69 & -0.01 \\ \cline{2-7}
                                                                           & \textbf{Snow}                               &  3.26 (1)  & 0.83  & \ 0.83  &  0.46  &  0.44  \\
\bottomrule
\end{tabular}
\end{specialtable}

\begin{paracol}{2}
\switchcolumn

With the aim to compare the SCIs values between snow and water, the average index value of a square of 10 $\times$ 10 pixels was extracted from the resulting index maps  and calculated for each of the two LCTs.
 The snow and water pixels’ were selected from large snow/water regions to guarantee the proper pixel designation in the Greenland region. The accurate location was verified by visually inspecting the area using the Landsat reference images shown in Figure~\ref{fig7} and the Google Earth $Pro^{TM}$ platform. Table~\ref{tab6} displays the mean SCIs values corresponding to snow and water, and Figure \ref{fig8} is the plot of the values of the Table \ref{tab6}.    

\begin{figure}[H]
\includegraphics[width=13 cm]{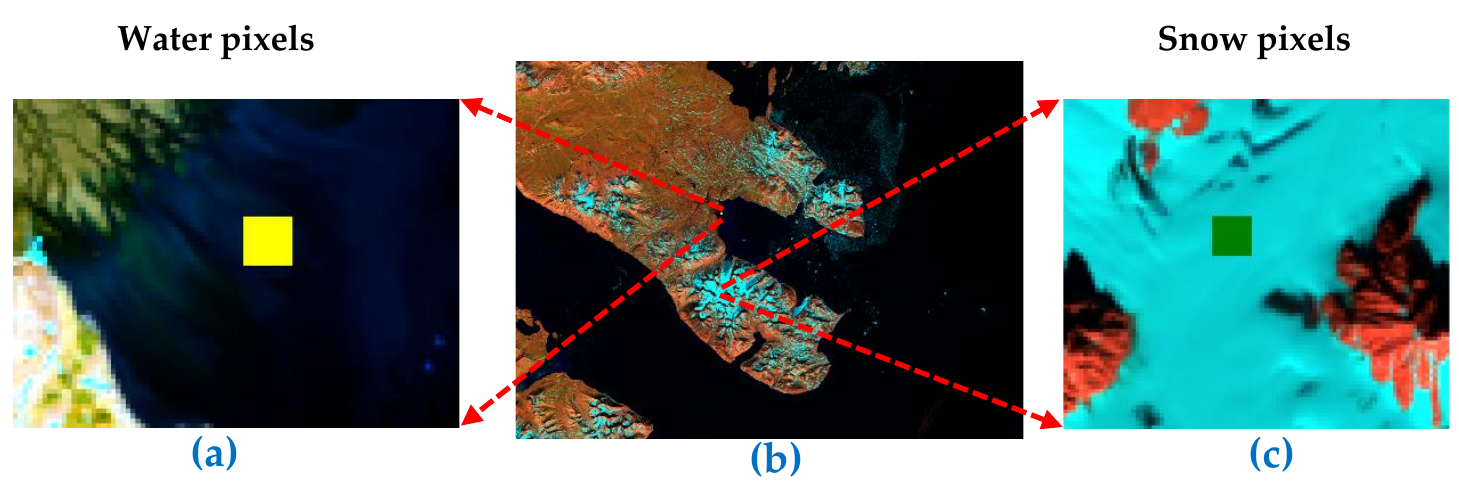}
\caption{Visualization of the   10 $\times$ 10 pixels squares taken in the Greenland region recorded by Landsat 8 OLI: (a) Sub-region of 10 $\times$ 10  water pixels (Yellow square), (b) Reference color image used to visualize  the amount of snow cover in blue color, and (c) Sub-region of 10 $\times$ 10 snow pixels (Green square). These squares are used to compare the capability of the Snow-Cover Indices maps in separating snow from water.\label{fig7}}
\end{figure}

Results in Table~\ref{tab6} and Figure \ref{fig8} show that NDSI, S3, and SWI methods present lower snow contrast compared to water, indicating the need for a water mask to remove it. Besides, the tied NDSII values between snow and water infer the need to identify an OTHV to improve its snow-cover map's delineation. On the other hand, the large contrast of NBSI-MS values between snow and water reaffirms no uncertainty in removing water in a NBSI-MS image.

\begin{specialtable} [H]
	\begin{minipage}{0.3\linewidth}
		\caption{Mean index values of one hundred pixels of water and snow taken from the Snow-Cover Indices maps on the Greenland region. Note: The NBSI-MS normalized values are in parenthesis.}
		\label{tab6}
		\centering
		\begin{tabular}{ccc}
\hline
\textbf{}        & \multicolumn{2}{c}{\textbf{Surface mean}} \\ \hline
SCIs             & \textbf{Water}       & \textbf{Snow}       \\ \hline
\textbf{SWI}     & 0.62                 & 0.48                \\ \hline
\textbf{S3}      & 0.66                 & 0.50                \\ \hline
\textbf{NDSII}   & 0.93                 & 0.94                \\ \hline
\textbf{NDSI}    & 0.94                 & 0.94                \\ \hline
\textbf{NBSI-MS} & -0.22 (-0.02)        & 8.07  (1)           \\ \hline
\end{tabular}
	\end{minipage} \hfill
	\begin{minipage}{0.45\linewidth}
		\centering
		\includegraphics[width=50mm]{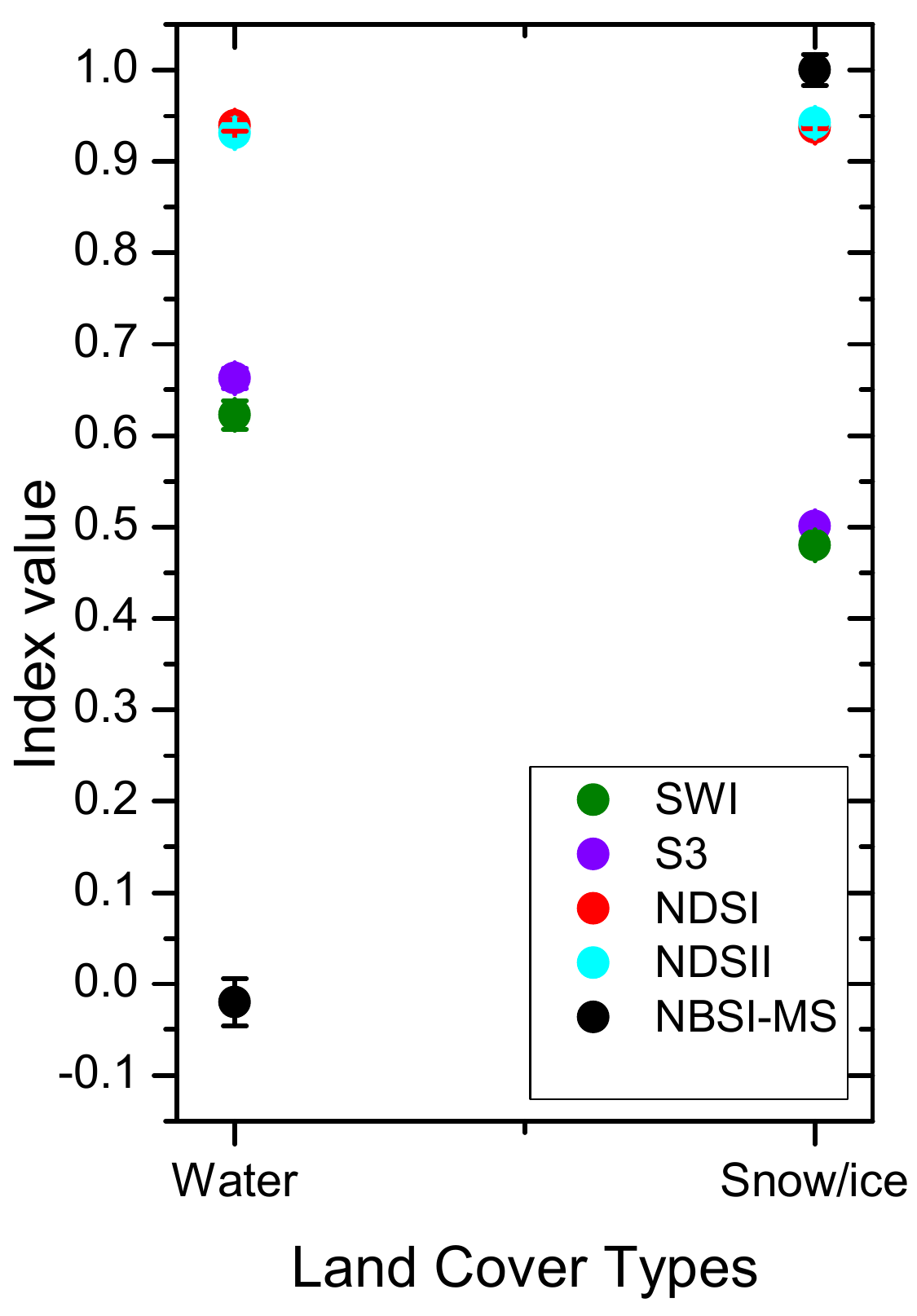}
		\captionof{figure}{Index values (mean $\pm$ standard deviation) of Table \ref{tab6}. The indices are ordered with increasing snow index values. }
		\label{fig8}

	\end{minipage}
\end{specialtable}

\subsection{Snow cover extraction maps} \label{sec:sec:one}

In this study, a qualitative comparison is made by visually inspecting the non-binary snow-cover extraction maps produced by the SCIs to identify their ability to delineate the snow-cover. Table~\ref{tab7} shows the description of the colored rectangles used to mark off the snow and non-snow spots presented in Figures~\ref{fig9} and \ref{fig10}.

\begin{specialtable}[H]
\centering
\caption{Description of the colored rectangles used to analyze the non-binary snow cover extraction maps
produced by the Snow-Cover Indices. \footnotesize{ Note: The color rectangles in the color composite images indicate the reference surface to inspect and rejection means a surface that the index algorithm has removed.} \label{tab7}}
\begin{tabular}{lll}
\toprule
\textbf{Images}                                      & \textbf{Color rectangle} & \textbf{Description} \\ \hline
\textbf{Reference color composite}  & Red                                           & Water surface                             \\ \cline{2-3}
                                                     & Green                                         & HS-V surface                              \\ \hline
\textbf{Snow cover extraction maps}& Blue                                          & Correctly rejected water surface          \\ \cline{2-3}
                                                     & Pink                                          & Misclassification of water surface        \\ \cline{2-3}
                                                     & White                                         & Correctly rejected HS-V                   \\ \cline{2-3}
                                                     & Yellow                                        & Misclassification of HS-V                 \\
\bottomrule
\end{tabular}
\end{specialtable}

Figure~\ref{fig9} shows the resultant snow cover extraction maps from the selected Greenland region produced by the SCIs. The NBSI-MS maps are visually compared against the NDSI, NDSII, S3, and SWI, taken as a reference to the false-color images. In the reference images, snow is blue,  water is black, and land is brown, while in the resultant snow extraction maps background should be black, and  snow is white or gray. The visual inspection where Landsat 5 TM and Landsat 8 OLI data are used shows that the NBSI-MS can reject the water surface with high performance, where the NDSI, NDSII-1, S3, and SWI methods fail to reject this kind of surfaces. Using Sentinel-2A MSI data, only the proposed NBSI-MS can reject the water surface correctly, where the other SCIs misclassified it. Similarly, in the France-Italy  region shown in Figure~\ref{fig10}, the proposed NBSI-MS method suppresses water surfaces with better precision than other SCIs, in all three datasets. Furthermore, the NBSI-MS method discriminates hilly shadows in vegetation (HS-V) whereas the rest of the SCIs fail to suppress them. The NDSI, NDSII, S3, and SWI did not remove the shadows in vegetation (HS-V) in the France-Italy scene registered by Landsat 8. Green, yellow and white rectangles represents this feature in the reference and the other images of  Landsat-8 in Figure \ref{fig10}. 


\end{paracol}
\nointerlineskip

\begin{figure}[H]
\centering
\includegraphics[width=16.5 cm]{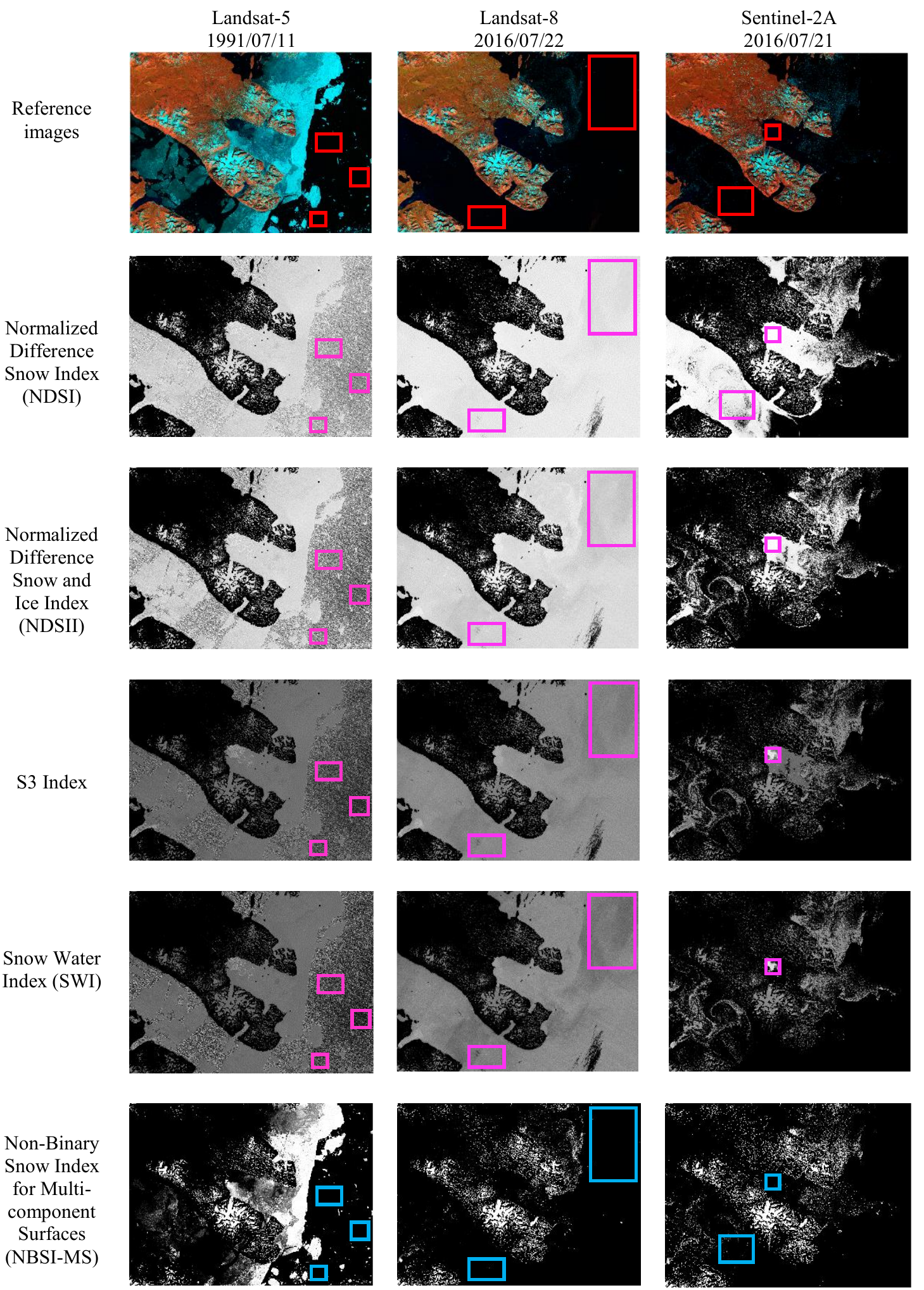}
\caption{Snow cover extraction maps produced by the Snow-Cover Indices without binarization in the Greenland region recorded by the Landsat and Sentinel-2A satellites on different dates. The reference  images are used to visualize  the amount of snow cover in blue color. \label{fig9}}
\end{figure}

\begin{figure}[H]
\centering
\includegraphics[width=13.7 cm]{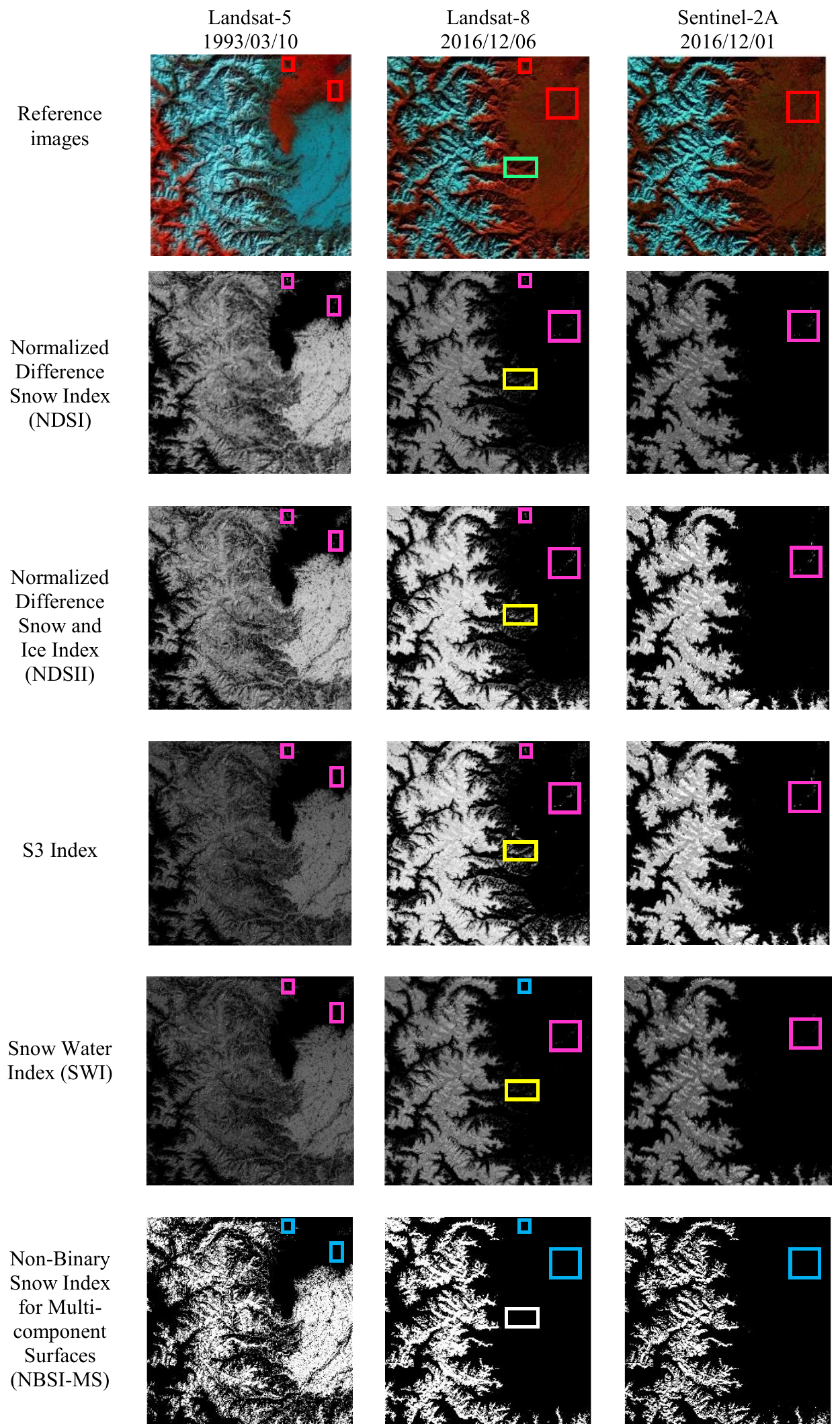}
\caption{Snow cover extraction maps produced by the Snow-Cover Indices without binarization in the France-Italy region recorded by the Landsat and Sentinel-2A satellites on different dates. The green rectangle in the resultant index maps of Landsat 8 highlights that NDSI, NDSII, S3, and SWI did not remove Hilly Shadow over Vegetation (yellow rectangles) as opposed to NBSI-MS (white rectangle). The reference color image are used to enhance  the amount of snow cover in blue color in the image.  \label{fig10}}
\end{figure}

\begin{paracol}{2}
\switchcolumn

 As a result, the NBSI-MS method can accurately delineate the snow/ice cover on non-binary maps in the Greenland and France-Italy regions recorded by the Landsat 5 TM, Landsat 8 OLI, and Sentinel-2A MSI satellites.
\par
According to  \cite{ref-Journal15}, SWI has shown better results than the NDSI, NDSII, and S3 methods for removing water pixels. Therefore, a thorough comparison between SWI and the proposed NBSI-MS methods is considered in this visual analysis. Figure~\ref{fig11} shows the non-binary snow-cover maps in the France-Italy region produced by SWI and the NBSI-MS methods. It can be seen that the NBSI-MS is capable of rejecting the water and HS-V surface with better performance compared with the SWI method.

\begin{figure}[H]
\centering
\includegraphics[width=13.7 cm]{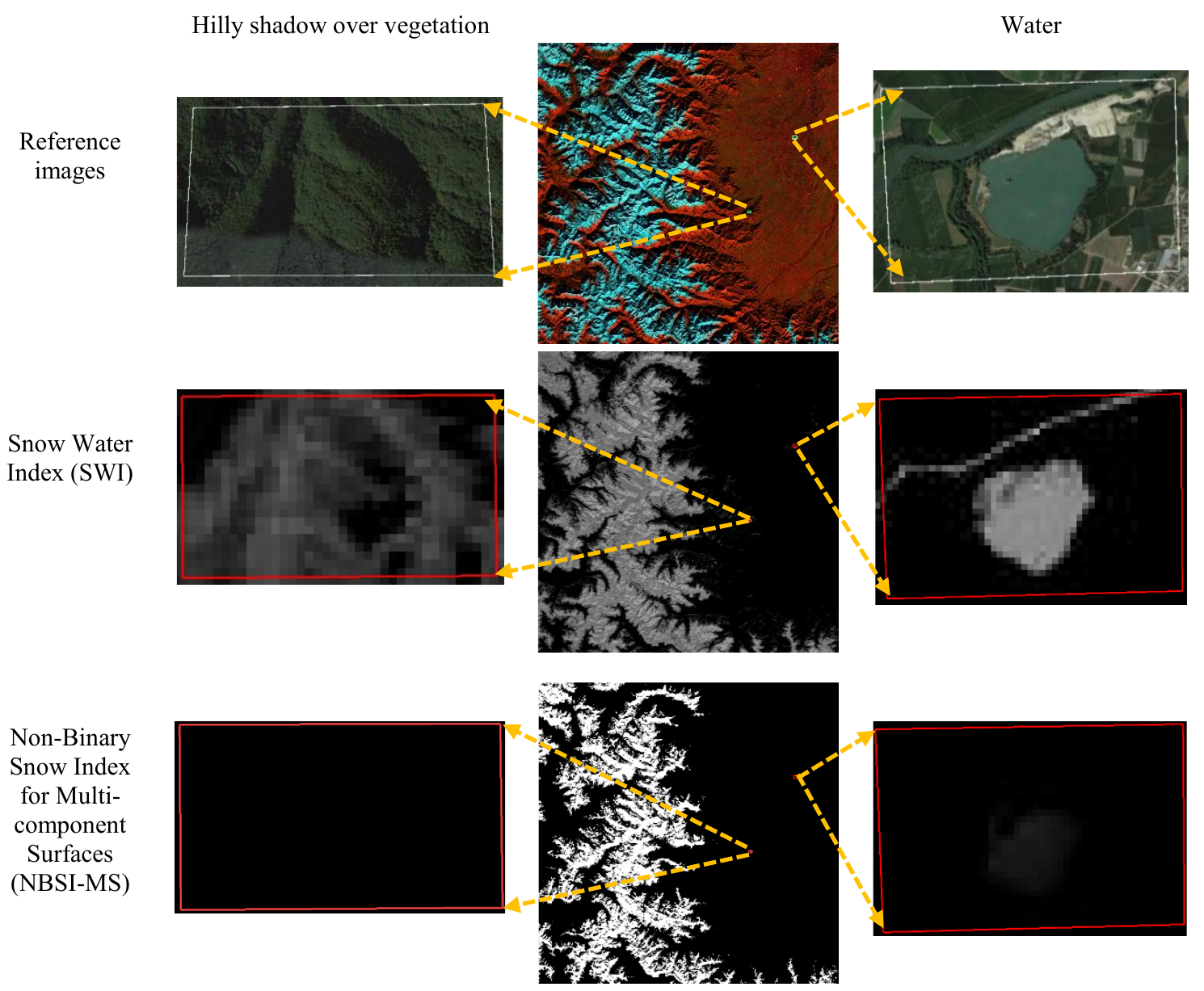}
\caption{ Comparison of performance for NBSI-MS and SWI methods. The reference images of Hilly Shadow over Vegetation (HS-V) and water are high resolution taken with Google Earth $Pro^{TM}$, while the center reference image is the false color used to visualize snow in blue. On the left column  the extracted HS-V areas are shown. The middle images are the resulting snow maps representing from where the areas were taken. The right side panels represent the extracted water areas. It can be seen that the NBSI-MS is capable of rejecting the water and HS-V surface with better performance compared with the SWI method since the images show dark results, while SWI presents gray pixels. 
\label{fig11}}
\end{figure}

\subsection{Precision assessment} \label{sec:sec:two} 

The SCIs must be evaluated based on the GRTPs reference data to quantitatively assess their efficiency in discriminating snow and non-snow pixels in the correct LCTs. In the virtue acquisition of the GRTPs reference data for evaluating the SCIs, over each scene, 150 GRTPs of snow and 150 GRTPs of non-snow were randomly generated. The 150 non-snow GRTPs were divided into six land cover types (Bare land, HS-BL, Impervious, Vegetation, HS-V, and water) in the France-Italy region. While in Greenland was divided into Bare land, vegetation, and HS-V as it only has 3 LCTs. The confusing random pixels on the edges were eliminated for the different LCTs by obtaining an equal amount for snow and non-snow pixels. To differentiate confusing snow and non-snow pixels, a high-resolution Google Earth $Pro^{TM}$ was used. The results are shown in Table \ref{tab8}. GRTPs using Landsat 5 TM data were generated separately from Landsat 8 OLI and Sentinel-2A MSI data to achieve the difference in snow-cover between scenes. 

\par

\begin{specialtable}[H]
\centering
\caption{Numbers of designated snow and non-snow Ground Reference Test Pixels (GRTPs) over Greenland and France-Italy regions. \label{tab8}}
\begin{tabular}{cccc}
\toprule
\multirow{2}{*}{\textbf{Region}}       & \multirow{2}{*}{\textbf{Satellite}} & \multicolumn{2}{c}{\textbf{GRTPs}} \\ \cline{3-4}
                                       &                                     & \textbf{Snow}  & \textbf{Non-snow} \\ \hline
\multirow{2}{*}{\textbf{Greenland}}    & Landsat 5 TM                        & 119            & 119               \\ \cline{2-4}
                                       & Landsat 8 OLI \& Sentinel 2A MSI    & 100            & 100               \\ \hline
\multirow{2}{*}{\textbf{Italy-France}} & Landsat 5 TM                        & 122            & 122               \\ \cline{2-4}
                                       & Landsat 8 OLI \& Sentinel 2A MSI    & 106            & 106               \\
\bottomrule
\end{tabular}
\end{specialtable}

The identification of mounting shadows in hilly areas was made by comparing the Landsat and Sentinel-2A data with the Digital Elevation Model (DEM) data,  downloaded from the web portal \url{https://search.earthdata.nasa.gov/search} \cite{ref-url5}. Figures~\ref{fig:fig12a} and \ref{fig:fig12b} depict the DEM in meters (m) of Greenland and France-Italy regions with snow and non-snow validation points.

\end{paracol}
\nointerlineskip

\begin{figure}[h!]
\captionsetup[subfigure]{justification=centering}
  \begin{subfigure}[b]{0.47\textwidth}
    \includegraphics[width=\textwidth]{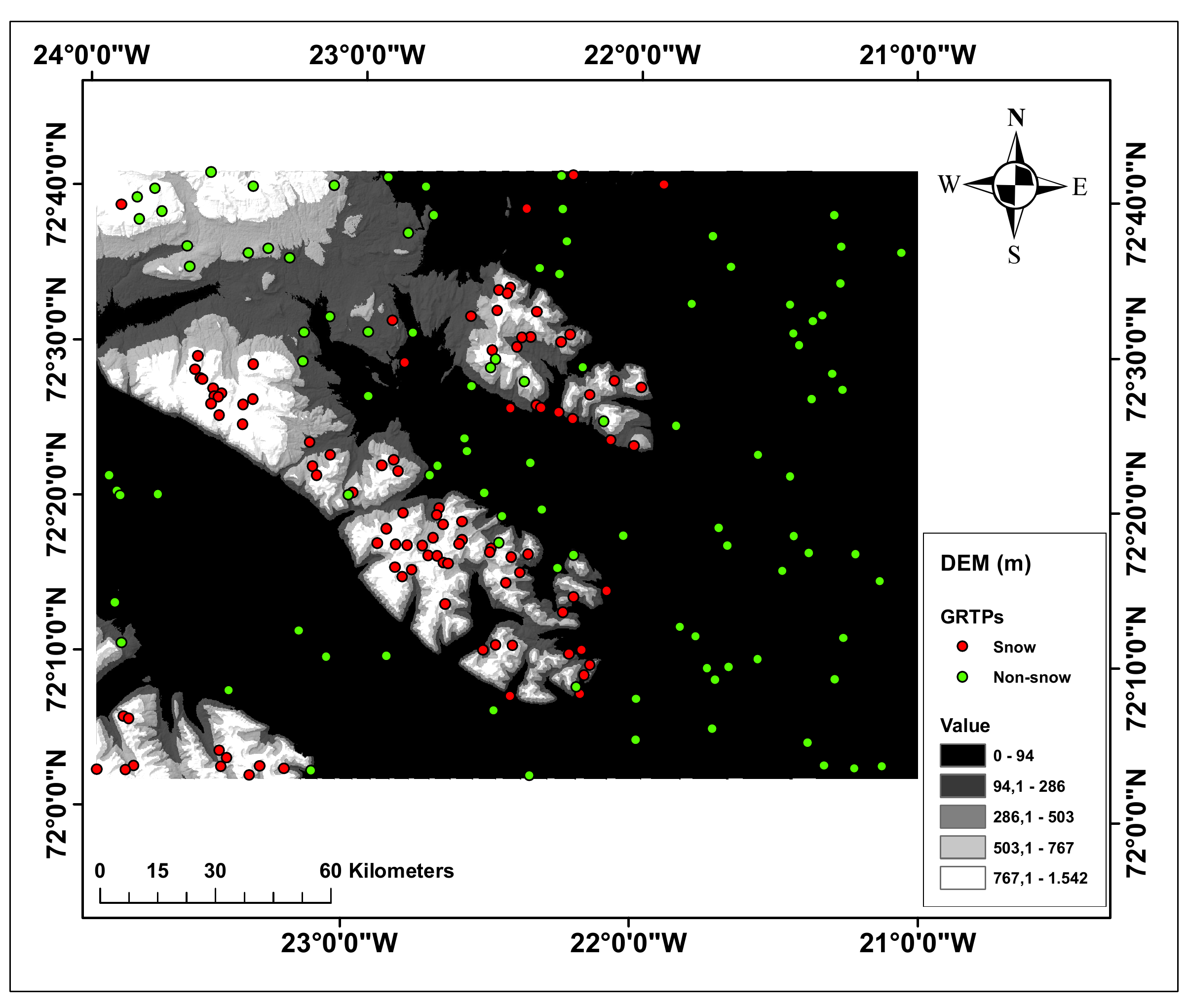}
    \caption{Greenland.}
    \label{fig:fig12a}
  \end{subfigure}
  \begin{subfigure}[b]{0.52\textwidth}
    \includegraphics[width=\textwidth]{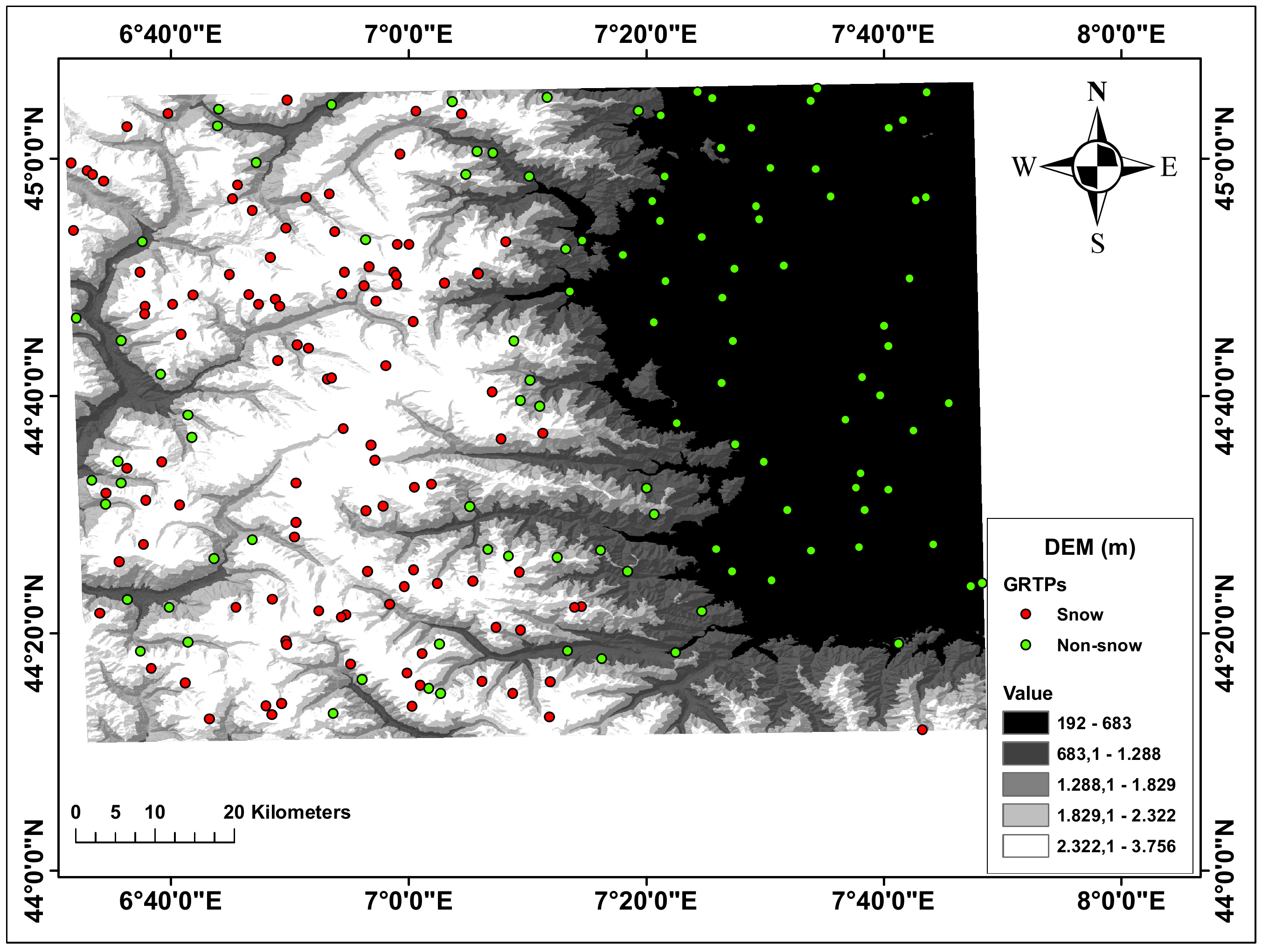}
    \caption{France-Italy .}
    \label{fig:fig12b}
  \end{subfigure}
  \caption{Distribution of the Ground Reference Test Pixels (GRTPs) over the Digital Elevation Model (DEM) images registered by the Advanced Spaceborne Thermal Emission and Reflection Radiometer (ASTER) of (a) The France-Italy and (b) The Greenland area. The DEM is used to verify the elevation ranges of the areas. \footnotesize{Note: The size of the GRTPs have been magnified to visualize the distribution. Moreover, the three red dots on the black area of the Figure \ref{fig:fig12a} are snow over islands and can be visualized in the reference images of Figure \ref{fig9}.}}
\end{figure}

\begin{paracol}{2}
\switchcolumn

For setting the OTHV, a range of separability between snow and non-snow index values must exist. Nevertheless, as shown in Table \ref{tab6}, the resulting mean of the water index values are larger than snow index ones for S3 and SWI, while for NDSI and NDSII are in the same range. Therefore, this study found that thresholding NDSI, S3, and SWI images for the Greenland region will remove most snow pixels. For this reason, all indices were evaluated using Eq. (\ref{eq:2}) to classify  snow and non-snow pixels.  
\par

The precision assessment of the SCIs was performed by comparing the snow/non-snow index extraction maps with the GRTPs. The purpose was to compute the number of: true positive (TP), false negative (FN), false positive (FP), and true negative (TN) pixels. Besides, these four-category pixel classifications were used to calculate the producer’s accuracy (PA), user’s accuracy (UA), overall accuracy (OA), and kappa coefficient \cite{ref-Journal36, ref-Journal37},  defined as:
 
\end{paracol}
\nointerlineskip
\begin{tabularx}{\textwidth}{XX}
\begin{equation}\label{eq:3}
    PA =  \frac{TP}{TP+FN} 
\end{equation}
    &
\begin{equation}\label{eq:4}
   OA = \frac{TP+TN}{P}
\end{equation}
\end{tabularx}\par

\begin{tabularx}{\textwidth}{XX}
\begin{equation}\label{eq:5}
    UA =  \frac{TP}{TP+FP}
\end{equation}
    &
\begin{equation}\label{eq:6}
   kappa = \frac{P(TP+TN)-\Sigma}{P^{2}-\Sigma}
\end{equation}
\end{tabularx}\par

\begin{paracol}{2}
\switchcolumn

Where {\em P} is the total number of the reference test pixels shown in Table ~\ref{tab8} and  $\Sigma$ is the chance accuracy and is calculated as:
\begin{equation} \label{eq:7}
\Sigma = (TP+FP)(TP+FN)+(FN+TN)(FP+TN)
\end{equation}

The PA, UA, and OA represent the accurate predictions ranging from 0 to 1, where 1 represents perfect accuracy. Nevertheless, these accurate predictions do not consider the agreements between datasets due to chance alone. The kappa coefficient typically ranges from -1 to 1, where 0 represents the agreement required from random chance, and 1 represents the absolute agreement between the raters \cite{ref-Journal36, ref-Journal37}. The precision assessment results of the SCIs are shown in Table~\ref{tab9}. It can be seen that the proposed NBSI-MS method has the highest OA and Kappa in a range of [0.99, 1]. Likewise, the quantitative evaluation confirms that the NBSI-MS method has high robustness for extracting snow/ice cover in data recorded by the Landsat 5 TM, Landsat 8 OLI, and Sentinel-2A MSI satellites.

\end{paracol}
\nointerlineskip

\begin{specialtable}[H]
\centering
\caption{Precision assessment based on Ground Reference Test Pixels validation for different Snow-Cover Indices. \label{tab9}}
\begin{tabular}{cccccccccc}
\toprule
\textbf{}                                  & \textbf{}                       & \multicolumn{4}{c}{\textbf{Greenland}}                                                                       & \multicolumn{4}{c}{\textbf{Italy-France}}                                                                               \\ \hline
\textbf{Satellite}                         & \textbf{Indices}                & \textbf{PA}               & \textbf{UA}               & \textbf{OA}               & \textbf{Kappa}            & \textbf{PA}                  & \textbf{UA}               & \textbf{OA}                   & \textbf{Kappa}                \\ \hline
                                           & NDSI                            & 1                         & 0.77                     & 0.85                     & 0.70                     & 1                            & 0.75                     & 0.84                         & 0.67                         \\ \cline{2-10} 
                                           & NDSII                           & 1                         & 0.77                     & 0.85                     & 0.70                     & 1                            & 0.76                     & 0.84                         & 0.69                         \\ \cline{2-10} 
                                           & S3                              & 1                         & 0.77                     & 0.85                     & 0.70                     & 1                            & 0.76                     & 0.84                         & 0.69                         \\ \cline{2-10} 
                                           & SWI                             & 1                         & 0.77                     & 0.85                     & 0.71                     & 1                            & 0.77                     & 0.85                         & 0.70                         \\ \cline{2-10} 
\multirow{-5}{*}{\textbf{Landsat 5 TM}}    & NBSI-MS & 1 & 1 & 1 & 1 & 1    & 1 & 1     & 1     \\ \hline
                                           & NDSI                            & 1                         & 0.55                     & 0.60                     & 0.20                      & 1                            & 0.74                     & 0.82                         & 0.64                         \\ \cline{2-10} 
                                           & NDSII                           & 1                         & 0.55                     & 0.59                      & 0.18                      & 1                            & 0.74                     & 0.83                         & 0.65                        \\ \cline{2-10} 
                                           & S3                              & 1                         & 0.56                     & 0.6                       & 0.2                       & 1                            & 0.75                     & 0.83                         & 0.66                         \\ \cline{2-10} 
                                           & SWI                             & 1                         & 0.56                     & 0.6                       & 0.2                       & 1                            & 0.84                     & 0.90                         & 0.80                        \\ \cline{2-10} 
\multirow{-5}{*}{\textbf{Landsat 8 OLI}}   & NBSI-MS & 1 & 1 & 1 & 1 & 0.99 & 1 & 1 & 1 \\ \hline
                                           & NDSI                            & 1                         & 0.83                     & 0.9                       & 0.8                       & 1                            & 0.9                     & 0.94                         & 0.89                         \\ \cline{2-10} 
                                           & NDSII                           & 1                         & 0.91                     & 0.95                      & 0.9                       & 1                            & 0.9                     & 0.94                         & 0.89                         \\ \cline{2-10} 
                                           & S3                              & 1                         & 0.91                     & 0.95                      & 0.9                       & 1                            & 0.90                     & 0.94                         & 0.89                         \\ \cline{2-10} 
                                           & SWI                             & 1                         & 0.95                     & 0.98                     & 0.95                      & 1                            & 0.91                     & 0.95                         & 0.9                         \\ \cline{2-10} 
\multirow{-5}{*}{\textbf{Sentinel-2A MSI}} & NBSI-MS                        &  1                         &  0.99                     &  1                     &  0.99                      &  1                            & 1                         &  1                             &  1                             \\ 
\bottomrule
\end{tabular}
\end{specialtable}

\begin{paracol}{2}
\switchcolumn


\section{Discussion} \label{sec:four}
In this study, the NBSI-MS method is proposed, and its performance for snow/ice cover mapping is compared with the well-known NDSI, NDSII, S3, and SWI methods. 
The analysis of the SCIs values in Table~\ref{tab5} showed that the NBSI-MS achieves considerable separability between snow and background, whereas the NDSI, NDSII, S3, and SWI deliver close index values between water, HS-V, and snow. To analyze the separation of the index values between snow and other LCTs the following equation was computed by using the SCIs values of Table~\ref{tab5},

\begin{equation}
\label{eq:8}
S = SIV - LCTIV,
\end{equation}
where \textit{S} is equal to separation value between snow and the different LCTs, SIV is the snow index value and LCTIV is the index value corresponding to each land cover type.  
\par
Table~\ref{tab10} shows that the NBSI-MS separation value between snow and bare land is $19.09$, while for the snow and water is $8.39$ using Landsat 8 OLI data. In Sentinel-2A data, the NBSI-MS separation value between snow and bare land is 12.58, and for snow and water is 3.60. On the other hand, it is noticed that NDSI, NDSII, S3, and SWI separation values corresponding to snow, HS-V, and water are close on both cases where Landsat 8 OLI and Sentinel-2A data are used.

\begin{specialtable}[H]
\centering
\caption{Snow-Cover Indices (SCIs) Separation values between snow and Land Cover Types (LCTs) in the Italy-France region using the Landsat 8 OLI and Sentinel-2A MSI datasets. \label{tab10}}
\begin{tabular}{ccccccc}
\toprule
                                 &                                            & \multicolumn{5}{c}{\textbf{\begin{tabular}[c]{@{}c@{}}Separation values between snow and  other \\ LCTs using the SCIs\end{tabular}} }
                                              \\ \cline{3-7} 
\multirow{-2}{*}{\textbf{Satellite}}      & \multirow{-2}{*}{\textbf{LCTs}} & \textbf{NBSI-MS} & \textbf{NDSI} & \textbf{NDSII} & \textbf{S3}                     & \textbf{SWI} \\ \hline
                                          & Bare land                             & 19.09          & 1.47        & 1.43         & 0.79                          & 0.58       \\ \cline{2-7} 
                                          & Impervious                            & 12.52          & 1.11        & 1.09         & 0.59                          & 0.59       \\ \cline{2-7} 
                                          & Vegetation                            & 11.45          & 1.26        & 1.45         & 0.72                          & 0.38       \\ \cline{2-7} 
                                          & Water                                 & 8.39           & 0.10        & 0.14         & 0.07                          & 0.22       \\ \cline{2-7} 
                                          & HS-V                                  & 9.37           & 0.06        & 0.19         & 0.10                          & 0.12       \\ \cline{2-7} 
\multirow{-6}{*}{\textbf{Landsat 8 OLI}}  & HS-BL                                 & 11.02          & 0.95        & 1.09         & 0.59                          & 0.59       \\ \hline
                                          & Bare land                             & 12.58          & 1.58        & 1.50         & 0.84                          & 0.60       \\ \cline{2-7} 
                                          & Impervious                            & 7.95           & 1.26        & 1.22          & 0.65                          & 0.68       \\ \cline{2-7} 
                                          & Vegetation                            & 9.50           & 1.62        & 1.76         & 0.88                          & 0.43       \\ \cline{2-7} 
                                          & Water                                 & 3.60           & 0.037        & 0.02         & -0.05 & 0.09       \\ \cline{2-7} 
                                          & HS-V                                  & 5.42           & 0.42        & 0.72         & 0.40                          & 0.39       \\ \cline{2-7} 
\multirow{-6}{*}{\textbf{Sentinel-2 MSI}} & HS-BL                                 & 5.42           & 1.82        & 1.77         & 1.15                          & 0.45       \\ 
\bottomrule
\end{tabular}
\end{specialtable}

According to the visual inspection of Figures~\ref{fig9} and \ref{fig10}, the worst performance of the NDSI, NDSII, S3, and SWI methods is exhibited by the Greenland region in the presence of a large water surface using Landsat 8 OLI data. Besides, the index values' separation between snow and water revealed that NDSI, S3, and SWI enhance water rather than snow, exhibiting negative values (Table~\ref{tab11}). This finding implies that these indices still need a water mask to improve their snow-cover delineation. In contrast, the NBSI-MS separation value between  snow and water is 8.29, indicating a net discriminative power.

\begin{specialtable}[H] 
\centering
\caption{Snow-Cover Indices separation values between snow and water in the Greenland region. \footnotesize{Note: The values were computed using Eq. (\ref{eq:8}) }\label{tab11}}
\begin{tabular}{cccccc}
\toprule
\textbf{} & \multicolumn{1}{c}{\textbf{NDSI}} & \multicolumn{1}{c}{\textbf{S3}}  & \multicolumn{1}{c}{\textbf{NDSII}} & \multicolumn{1}{c}{\textbf{SWI}} & \multicolumn{1}{c}{\textbf{NBSI-MS}} \\ \hline
\textbf{Snow-water}               & 0.00  & -0.16 & 0.01    & -0.14 & 8.29      \\ 
\bottomrule
\end{tabular}
\end{specialtable}

This work addresses the important issue for the remote sensing community of developing  higher and higher accurate methods to identify local areas from remotely sensed images. This could be possible by collecting higher-resolution images with newer satellite technologies in smaller areas. Therefore, while we have indexes (like the traditional NDSI) with the ability to analyse large (500-m) areas, the proposed non-binary multispectral indexes should be used to provide complementary information scaling down the investigated area sizes.  Further developments of this work have been devised to create a hierarchical structure of interrelated nested indexes within areas at different scales. 
\par
Several studies have addressed the issue of integrating high resolution pixel data from  different bands in low resolution pixel data. For example,  the NDSI snow index extracted from Landsat 30-m  was scaled up and compared to the MODIS 500-m pixel based on a linear regression approach in \cite{salomonson2004estimating,wang2020universal}. Fully grasping  and exploiting sub-pixel information content while tackling the computational complexities of dealing with huge amount of data on a bidaily, automated, global basis still pose a challenge. However, more realistic applications of the multiple-resolution approach with the integration of fractional and binary snow-cover data could be devised, as: (1)  analyzing  snow-cover metamorphism to complement local-field measurement of mechanical properties of snow to increase the ability of trigger real-time alert in case of adverse meteorological event condition \cite{carbone2010snow}; (2)  providing sub-pixel information for calibrating or verifying hydrological models at small and intermediate scales \cite{zhang2021enhanced}.
\par
The widely used binary (i.e., snow or non-snow) index  data uses the assumption that above the threshold the pixel is covered by snow. However the spatially fixed  threshold, might not be optimal for local applications with variations in landscape and satellite viewing conditions \cite{ref-Journal24}. 
Further developments of this research have been devised to implement multiple resolution estimate in terms of a pixel-by-pixel scale dependent  approach (as in \cite{valdiviezo2014hurst})  rather than the simple linear regression.

\section{Conclusions} \label{sec:five}

The main objective of this research is to present the NBSI-MS method and compare its capability for mapping the snow/ice cover against the NDSI, NDSII, S3, and SWI methods in the presence of vegetation, water, impervious, bare land, HS-V, HS -BL, and snow/ice. The analysis of all indices was done in the same image conditions according to the pre-processing steps in non-binarized results using Landsat 5, Landsat 8, and Sentinel-2A scenes. Image thresholding or the application of masking techniques have not been implemented in this analysis since the proposed NBSI-MS method does not need any additional techniques to delineate the snow/ice with high accuracy in different environmental conditions.
\par
The NBSI-MS method shows a strong potential for snow cover mapping. Qualitative and quantitative results of this research have shown that the NBSI-MS method has a higher accuracy over NDSI, NDSII, S3, and SWI methods. Furthermore, the NBSI-MS values confirm that non-normalized index increases the contrast between snow and background, providing high-quality delineation of snow/ice cover on non-binary maps using Landsat and Sentinel-2A data.
The  most outstanding results of the comparison among the SCIs are: 

\begin{itemize}
	\item Figures~\ref{fig:fig6a}, \ref{fig:fig6b}, and Table~\ref{tab5} reveal that the proposed NBSI-MS method discriminates the background as bare land, impervious, vegetation, water, HS-V, and HS-BL with index values below zero. In contrast, the snow/ice is highlighted with positive NBSI-MS values. On the other hand, NDSI, NDSII, S3, and SWI deliver positive index values on water, and HS-V closes to their snow index values. 
	
	\item Table~\ref{tab6} shows  negative NBSI-MS values for water, while snow has positive NBSI-MS value, which confirms a large contrast between them. Conversely, NDSI, S3, and SWI highlight water over snow, and the close NDSII values between water and snow indicate the need for a water mask to enhance snow-cover maps in the Greenland region.
	
	\item The visual inspection presented in Figures~\ref{fig9} and \ref{fig10} show that NBSI-MS can reject the water and HS-V surfaces correctly, while NDSI, NDSII, S3, and SWI fail to suppress them. Besides, the NDSI, NDSII, S3, and SWI perform better in the France-Italy region, but they presented a poor performance in the Greenland region in the presence of a large water surface.
	
	\item The precision assessment results of Table~\ref{tab9} show that the SCIs reach a high PA, which means high precision for extracting the snow-pixels. However, in the Greenland region using Landsat data, the NDSI, NDSII, S3, and SWI methods show low UA, which means misclassification of non-snow pixels. Whereas, the proposed NBSI-MS method has the highest OA and Kappa in the range [0.99, 1] in Greenland and France-Italy regions, demonstrating higher precision than NDSI, NDSII, S3, and SWI methods. 
	
	\item  As shown in Figure \ref{fig4}, pre-processing steps must be taken into account to obtain high-quality NBSI-MS maps since the algorithm has shown greater sensitivity to atmospheric conditions than the compared indices. 
\end{itemize}
In summary, the very good agreement between qualitative and quantitative results confirms the performance superiority of the proposed NBSI-MS method for removing water and shadow pixels over NDSI, NDSII, S3, and SWI methods. The NBSI-MS values confirm that the high selectivity of the index and its ability to discriminate between snow and background, providing high-quality delineation of snow/ice cover on non-binary maps of Landsat and Sentinel-2A data.


\vspace{6pt} 



\authorcontributions{Investigation, M.M.A., M.D., C.T., A.P., J.G.O., and A.C.; Conceptualization, M.M.A., M.D., C.T., A.P., J.G.O., and A.C.; Methodology M.M.A., and M.D.; software, M.M.A., M.D., C.T., A.P., J.G.O., and A.C.; validation M.M.A.; Formal analysis, M.M.A., and M.D.; Funding acquisition, J.G.O., C.T. A.P., and A.C.; Project coordination, A.C.,and C.T.; Writing—original draft, M.M.A., M.D., C.T., A.P., A.C., and J.G.O. All authors have read and agreed to the published version of the manuscript.}

\funding{This work was supported by the Consejo Nacional de Ciencia y Tecnología (CONACyT) (scholarships 858585 and 858588), the European Commission (Grant JTC-2016-004 FlagERA - FuturICT2.0) and the Politecnico di Torino (Fondi di Ateneo - Supporto Ricerca di Base).}

\acknowledgments{M. Arreola-Esquivel and M. Delgadillo-Herrera with 858585 and 858588 CVUs are thankful to Consejo Nacional de Ciencia y Tecnología (CONACyT) for the scholarship awarded. Also, M.M.A. and M.D. acknowledge financial support during their research stay at the Politecnico di Torino in October-November 2018 and May-June 2019 (Grant JTC-2016-004 Flagera FuturICT2.0).}

\conflictsofinterest{The authors declare no conflict of interest.} 





\end{paracol}
\newpage
\reftitle{References}

\end{document}